\documentclass[prx,twocolumn,superscriptaddress]{revtex4-1}
\usepackage{latexsym,bm}
\usepackage{amsmath}
\usepackage{graphicx}
\usepackage{subfigure}

\begin{document}


\title{Depict noise-driven nonlinear dynamic networks from output data by using high-order correlations}



\author{Chen Yang}
\affiliation{School of Sciences, Beijing University of Posts and Telecommunications - Beijing, China}

\author{Zhang Zhaoyang}
\affiliation{Faculty of Science, Ningbo University - Ningbo, China}

\author{Chen Tianyu}
\affiliation{School of Sciences, Beijing University of Posts and Telecommunications - Beijing, China}

\author{
Wang Shihong$\ ^{1,\,\dag}$
}
\noaffiliation
\affiliation{School of Sciences, Beijing University of Posts and Telecommunications - Beijing, China}

\author{Hu Gang} 
\email[]{
ganghu@bnu.edu.cn;\  $^\dag$\ \  shwang@bupt.edu.cn}
\affiliation{Department of Physics, Beijing Normal University - Beijing, China}

%
%


\date{\today}

\begin{abstract}
  Many practical systems can be described by dynamic networks,
  for which modern technique can measure their output signals, and accumulate extremely rich data.
  Nevertheless, the network structures producing these data are often deeply hidden in these data.
  Depicting network structures by analysing the available data, i.e.,
  the inverse problems, turns to be of great significance.
  On one hand, dynamics are often driven by various unknown facts, called noises.
  On the other hand, network structures of practical systems are commonly nonlinear,
  and different nonlinearities can provide rich dynamic features and meaningful functions of realistic networks.
  So far, no method, both theoretically or numerically, has been found to systematically treat both difficulties together.
  Here we propose to use high-order correlation computations (HOCC) to treat nonlinear dynamics;
  use two-time correlations to treat noise effects;
  and use suitable basis and correlator vectors to unifiedly depict all dynamic nonlinearities, topological interaction links and noise statistical structures.
  All the above theoretical frameworks are constructed in a closed form and numerical simulations fully verify the validity of theoretical predictions.

\end{abstract}


\maketitle

  \section{introduction}
  In recent decades, the topic of dynamical complex networks has attracted great attention in interdisciplinary fields due to its theoretical importance and practical significance \cite{N1,N2}.
  It is well aware that network dynamic structures is determined in great extent by network structures, mainly classified by dynamics of local nodes and interactions between the nodes of networks.
  In practical cases, we can measure the outputs of network nodes while the structure of networks are often deeply hidden in the measured data.
  Therefore, it turns to be crucial to develop effective methods to depict network structures from the available data of nodes.
  This is the so-called inverse problem, which  becomes one of the most important topics in the data analysis of complex networks in wide crossing fields, in particular, in biology and social science \cite{N3,N4,N5,N6,N7}.

  Various methods have been proposed to treat the inverse problems \cite{RA}.
  There are several typical difficulties in practice.
  First, in most of realistic cases diverse facts, termed as noises, are involved in the data production.
  These noises make the structure depiction difficult because they are unknown on one hand,
  and essentially influence the data analysis on the other hand.
  Different statistical methods based on various correlation computations have been suggested to treat the noise problems \cite{N9,N9-3,N10,N13,N15,N17}.
  Second, in almost all realistic network systems various nonlinearities play crucial roles in generating diverse characteristic features and significant functions.
  However, so far all works in treating network inference have made approximations either neglecting noise influences \cite{N8,N9-1,N11,N12,N14,N16},
  or considering linear dynamics and interactions (or linearized these items) \cite{N9,N9-3,N10,N13,N15,N17}.
  No one considered the two facts jointly.
  These methods fail when both noise effects and nonlinearities of network structures are crucial for the data production.

  In this presentation we consider the inverse problem of noise-driven nonlinear dynamic networks.
  The key point in dealing with this problem is:
  We compute high-order correlations to treat possible nonlinear structures,
  and with the help of these correlations we convert the statistics of inference of noise-driven nonlinear networks to linear matrix algebraic computations.
  In the next section the idea and the method how to use high-order correlation computation (HOCC) to infer nonlinear dynamic networks are explained,
  including to depict all the internal node dynamics, mutual interactions and correlation structures of multiplicative noises.
  In Sec.III we first apply the HOCC method to a simple three node network,
  the well known Lorenz equation,
  driven by white noises.
  And then large complex networks, with diverse nonlinearities in local dynamics;
  complicated links between nodes; and diverse noise correlations, are considered.
  The effectiveness of the HOCC algorithm are again well justified.
  In Sec.IV networks with complicated nonlinear phase dynamics, nonlinear interactions and nonlinear multiplicative noise statistics are investigated.
  The HOCC method with properly chosen basis vectors and high-order correlation types are applied to satisfactorily overcome all the difficulty by there diverse complexities.
  Section V gives some conclusion and perspective of the applications of the method in practical inverse problems. \vfill
  
  \section{Inferring nonlinear networks by using high-order and two-time-point correlations: theory}
  Let us consider a very general noise-driven nonlinear network
  \begin{eqnarray}
    \dot{\bm{x}}(t) & = & \bm{f}(\bm{x}(t))+\bm{\Gamma}(t) \label{GeneralEq} \\
    \bm{x} & = & (x_1,x_2,\cdots,x_N)^T \nonumber \\
    \bm{f} & = & (f_1,f_2,\cdots,f_N)^T \nonumber \\
    \bm{\Gamma} & = & (\Gamma_1,\Gamma_2,\cdots,\Gamma_N)^T \nonumber
  \end{eqnarray}
  where noises $\Gamma_i(t)$, $i=1,2,\cdots,N$, represent impacts from microscopic world,
  and they are expected to have very short correlation time $t_{cor}\ll1$,
  much smaller than the characteristic time of deterministic dynamics assumed to be of order 1.
  Then noises are approximated as white ones,
  \begin{equation}
  <\Gamma_i(t)>=0,\ <\Gamma_i(t)\Gamma_j(t^\prime)>=Q_{ij}(\bm{x})\delta(t-t^\prime) \label{WN}
  \end{equation}
  with $i,j=1,2,\cdots,N$.
  Here noises are sufficiently strong so that any deterministic solutions are impossible \cite{N12,N9-1}.
  It is emphasized that this is the first time to consider multiplicative noises in the study of inverse problems due to its essential difficulty,
  though this type of noises exist extensively in practical circumstances \cite{N20,N21,N22,N23}.
  In Eq.(\ref{GeneralEq}) we have all measurable data in our hand,
  namely, we measure
  \begin{gather}
    \bm{x}(t_1),\bm{x}(t_2),\cdots,\bm{x}(t_k),\cdots,\bm{x}(t_L) \label{LTS}\\
    t_{cor}<\Delta t=t_{k+1}-t_k\ll 1;\ k=1,2,\cdots,L;\ L\gg1  \nonumber
  \end{gather}
  With $\Delta t\ll1$ we can compute velocities of $\bm{x}$ in Eq.(\ref{GeneralEq}) and with $L\gg1$ we have sufficiently large samples to perform statistics.
  These conditions are not always available, but they do be available in many important practical experiments,
  or can be realized on purpose in case of need.

  In Eq.(\ref{GeneralEq}) all functions $f_i$, $i=1,2,\cdots,N$ are unknown.
  The noise statistic matrix $\hat{\bm{Q}}(\bm{x})=(Q_{ij}(\bm{x}))$ are unknown either.
  Only the output variables (\ref{LTS}) are available for analysis.
  The task is to depict dynamic functions $f_i$, and noise statistics $Q_{ij}$, $i,j=1,\cdots,N$, including their nonlinearities of node dynamics and interactions between nodes.
  This task can be fulfilled either without noise $Q_{ij}=0$, $i,j=1,\cdots,N$ \cite{N8,N9-1,N11,N12,N14,N16}
  or with field $\bm{f}=(f_1,\cdots,f_N)^T$ linear \cite{N9,N9-3,N10,N13,N15,N17}.
  With both strong noises and essential nonlinearities effective in Eq.(\ref{GeneralEq}),
  there has been no method making network structure depiction, and this is the task of the present work.

  For start we assume $f_i$s can be very generally expanded by a certain basis set \cite{N9-1,N9-2}
  \begin{equation}
    f_i(\bm{x}) = \sum_{\mu=1}^\infty A_{i,\mu} Y_{i,\mu}(\bm{x}),\ i=1,2,\cdots,N \label{GExp}
  \end{equation}
  where all constant coefficients $A_{i,\mu}$, $\mu=1,2,\cdots,M_i,\cdots,\infty$ are unknown,
  while all functions $Y_{i,\mu}(\bm{x})$ called as bases are known.
  For treating nonlinearities in $\bm{f}(\bm{x})$, the chosen basis set should be complete for expanding field $\bm{f}(\bm{x})$.
  In Eq.(\ref{GeneralEq}) we give a freedom to use different basis sets suitable for expanding different field $\bm{f}(\bm{x})$.
  It seems that the expansions of Eq.(\ref{GExp}) should include infinite terms for arbitrary functions $\bm{f}(\bm{x})$.
  One has to truncate the expansion to finite terms.
  There is a systematical and self-consistent method, described in following section to make such truncation.
  At the present, we just assume a truncation at $M_i$ for $f_i(\bm{x})$ expansion.
  Then Eq.(\ref{GExp}) can be simplified as
  \begin{eqnarray}
    f_i(\bm{x}(t)) & = & \bm{A}_i\bm{Y}_i(t) \label{GEqMat} \\
    \bm{A}_i & = & (A_{i,1},A_{i,2},\cdots,A_{i,M_i}) \nonumber \\
    \bm{Y}^T_i & = & (Y_{i,1}(t),Y_{i,2}(t),\cdots,Y_{i,M_i}(t)) \nonumber
  \end{eqnarray}

  Without noise $\Gamma_i(t)$, all the unknown coefficients can be solved by algebraic equations if sufficient data are accumulated \cite{N9-1,N12}.
  With strong noises the inference computations are much more difficult.
  Here for network depiction we use a method to compute two-time correlations to filter noise effect \cite{N13,N15},
  together with using high-order correlations on the chosen basis set to treat nonlinearities of the networks and multiplicative noises for depiction.

  For arbitrary node $i$ in the network, multiplying Eq.(\ref{GeneralEq}) from the right side by a correlator vector $\bm{Z}_i^T(\bm{x}(t-\tau))$
  \begin{equation}
    \bm{Z}_i(\bm{x}) = (Z_{i,1}(\bm{x}),Z_{i,2}(\bm{x}),\cdots,Z_{i,M_i}(\bm{x}))^T \label{Correlor}
  \end{equation}
  and computing all related correlations,
  we obtain a linear matrix algebraic equation
  \begin{equation}
    \bm{B}_i(-\tau)=\bm{A}_i\hat{\bm{C}}_i(-\tau)+<\Gamma_i(t)\bm{Z}_i^T(t-\tau)> \label{BCG}
  \end{equation}
  with
  \begin{alignat}{2}
    \bm{B}_i(-\tau) &= (B_{i,1}(-\tau),B_{i,2}(-\tau),\cdots,B_{i,M_i}(-\tau)) \nonumber \\
    B_{i,\mu}(-\tau) &= <\dot{x}_i(t)Z_{i,\mu}(t-\tau)> \nonumber \\
        &= \frac{1}{L-p}\sum_{k=p+1}^L\dot{x}_i(t_k)Z_{i,\mu}(t_{k-p}) \label{MatDef} \\
    \dot{x}_i(t_k) &= \frac{x(t_{k+1})-x(t_k)}{\Delta t},\  \tau = p\Delta t \nonumber\\
    \hat{\bm{C}}_i(-\tau) &= (C_{i,\mu\nu}) \nonumber\\
        &= (\frac{1}{L-p}\sum_{k=p+1}^L Y_{i,\mu}(t_k)Z_{i,\nu}(t_{k-p})) \nonumber
  \end{alignat}
  All correlators $\bm{Z}_1(\bm{x}(t))$,$\bm{Z}_2(\bm{x}(t))$,$\cdots$,$\bm{Z}_N(\bm{x}(t))$ can be arbitrarily chosen under the condition that their entries must not be linearly dependent on each other so that matrix $\hat{\bm{C}}_i$ has full rank,
  and is invertible.
  In Eq.(\ref{MatDef}) we should have $t_{cor}<\tau\ll1$ with $\tau$ being larger than the correlation time of noises,
  and much smaller than the characteristic times of deterministic network dynamics, previously assumed to be of order 1.
  Now the noise-correlator correlation must vanish
  \begin{equation}
    <\Gamma_i(t)\bm{Z}_i(t-\tau)^T>\approx0,\  \mbox{for}\  \tau>t_{cor} \label{Decorrelation}
  \end{equation}
  since the fast-varying noises of Eq.(\ref{WN}) have no correlation with any variable data of earlier times,
  disregarding any forms of multiplicative noises $Q_{ij}(\bm{x})$.
  Now with the noise-decorrelation of Eq.(\ref{Decorrelation}), Eq.(\ref{BCG}) can be reduced to
  \begin{equation}
    \bm{B}_i(-\tau)=\bm{A}_i\hat{\bm{C}}_i
  \end{equation}
  leading to
  \begin{equation}
    \bm{A}_i = \bm{B}_i(-\tau)\hat{\bm{C}}_i^{-1}  \label{Outcome1}
  \end{equation}
  and with $\bm{B}_i(-\tau)$, $\hat{\bm{C}}_i$ and $\bm{A}_i$ being given in Eqs.(\ref{MatDef}) and (\ref{GEqMat}), respectively.
  We delete the notion $(-\tau)$ in $\hat{\bm{C}}_i$ because $\tau\ll1$ does not considerably change the values of $C_{i,\mu\nu}$.
  All elements of vector $\bm{B}_i(-\tau)$ and matrix $\hat{\bm{C}}_i$ can be computed with known output variables $\bm{x}(t)$,
  and thus all the unknown linear and nonlinear coefficients in Eq.(\ref{GeneralEq}) can be depicted,
  though the noise statistics $\hat{\bm{Q}}$ for Eq.(\ref{WN}) is unknown.

  Statistical features of noises play crucial roles in the data production.
  Depicting noise statistics is also important to understand the nature of data and to predict future data production in practical cases.
  It is desirable that noise matrix $\hat{\bm{Q}}(\bm{x})$ can be also easily depicted from the variable data by the a HOCC algorithm similar to the above
  \begin{eqnarray}
    Q_{ij}(\bm{x}) & = & \sum_{\mu=1}^{M_{ij}}D_{ij,\mu}q_{ij,\mu}(\bm{x}) \nonumber \\
    \bm{D}_{ij} & = & \Delta \bm{B}_{ij}\hat{\bm{G}}_{ij}^{-1}  \label{QEq}
  \end{eqnarray}
  where
  \begin{eqnarray}
    \Delta B_{ij,\nu}(\tau) & = & <\dot{x}_i(t)(x_j(t+\tau)-x_j(t-\tau))q^\prime_{ij,\nu}(\bm{x}(t-\tau))> \nonumber \\
     & = & \frac{1}{L-2p}\sum_{k=p+1}^{L-p}\dot{x}_i(t_k)(x_j(t_{k+p})-x_j(t_{k-p}))\cdot \nonumber \\
     &  & q^\prime_{ij,\nu}(\bm{x}(t_{k-p}))  \nonumber \\
    G_{ij,\mu\nu} & = & \frac{1}{L-p}\sum_{k=1+p}^L q_{ij,\mu}(\bm{x}(t_{k}))q^\prime_{ij,\nu}(\bm{x}(t_{k-p}))  \label{QEdef}
  \end{eqnarray}
  where $q_{ij,\mu}(\bm{x})$ are known bases for expanding $Q_{ij}(\bm{x})$;
  $G_{ij,\mu\nu}$ are matrix elements computable from correlations of bases $q_{ij,\mu}$ and corresponding correlator $q^\prime_{ij,\nu}$;
  and $D_{ij,\mu}$ are unknown coefficients to be depicted.
  The detailed derivation of Eqs(\ref{QEq}) and (\ref{QEdef}) are presented in Appendix.

  For numerical simulation we can arbitrarily choose the correlator vectors and we in the following computations simply take,
  \begin{eqnarray}
    Z_{i,\mu} & = & Y_{i,\mu},\ i=1,2,\cdots,N;\mu=1,2,\cdots,M_i \label{CorrelorChoice1} \\
    q^\prime_{ij,\mu} & = & q_{ij,\mu},\ i,j=1,2,\cdots,N;\mu=1,2,\cdots,M_{ij} \nonumber
  \end{eqnarray}
  which can easily guarantee the invertibility of $\hat{\bm{C}}_i$ and $\hat{\bm{G}}_{ij}$.

  Now four novel points of the present method should be emphasized.
  (i) High-order correlations are used to depict nonlinear structure of networks without pursuing any linearization approximation;
  (ii) Two-time-point correlations have been proposed to treat noise-decorrelation, and the time difference $\tau$ can be adjusted to suit different noise conditions;
  (iii) We suggested to freely design basis set $\bm{Y}_i,\bm{q}_{ij}$ and correlator set $\bm{Z}_i,\bm{q}^\prime_{ij}$ to construct correlation matrices under the condition of invertibility.
  (iv) For the first time, multiplicative noises are taken into account in the study of inverse problems.
  Multiplicative factors can be inferred together with the nonlinear fields and interaction structures.
  In Eq.(\ref{QEq})(\ref{QEdef}) $q_{ij,\mu}(\bm{x})$ and $q_{ij,\mu}^\prime(\bm{x})$, $\mu=1,2,\cdots,M_{ij}$, play roles of vector bases
  for expanding multiplicative factors $Q_{ij}(\bm{x})$,
  just like bases $\bm{Y}_i(\bm{x})$ and $\bm{Z}_i(\bm{x})$ for computing fields $f_i(\bm{x})$ in Eq.(\ref{GEqMat})(\ref{Correlor}).
  And $M_{ij}$ is the truncation in $Q_{ij}(\bm{x})$ expansion like $M_i$ for $f_i(\bm{x})$.
  \begin{figure}
  \begin{minipage}[c]{0.48\columnwidth}
    \centering
    \subfigure{\includegraphics[angle=0,width=0.96\textwidth,totalheight=0.5\textwidth]{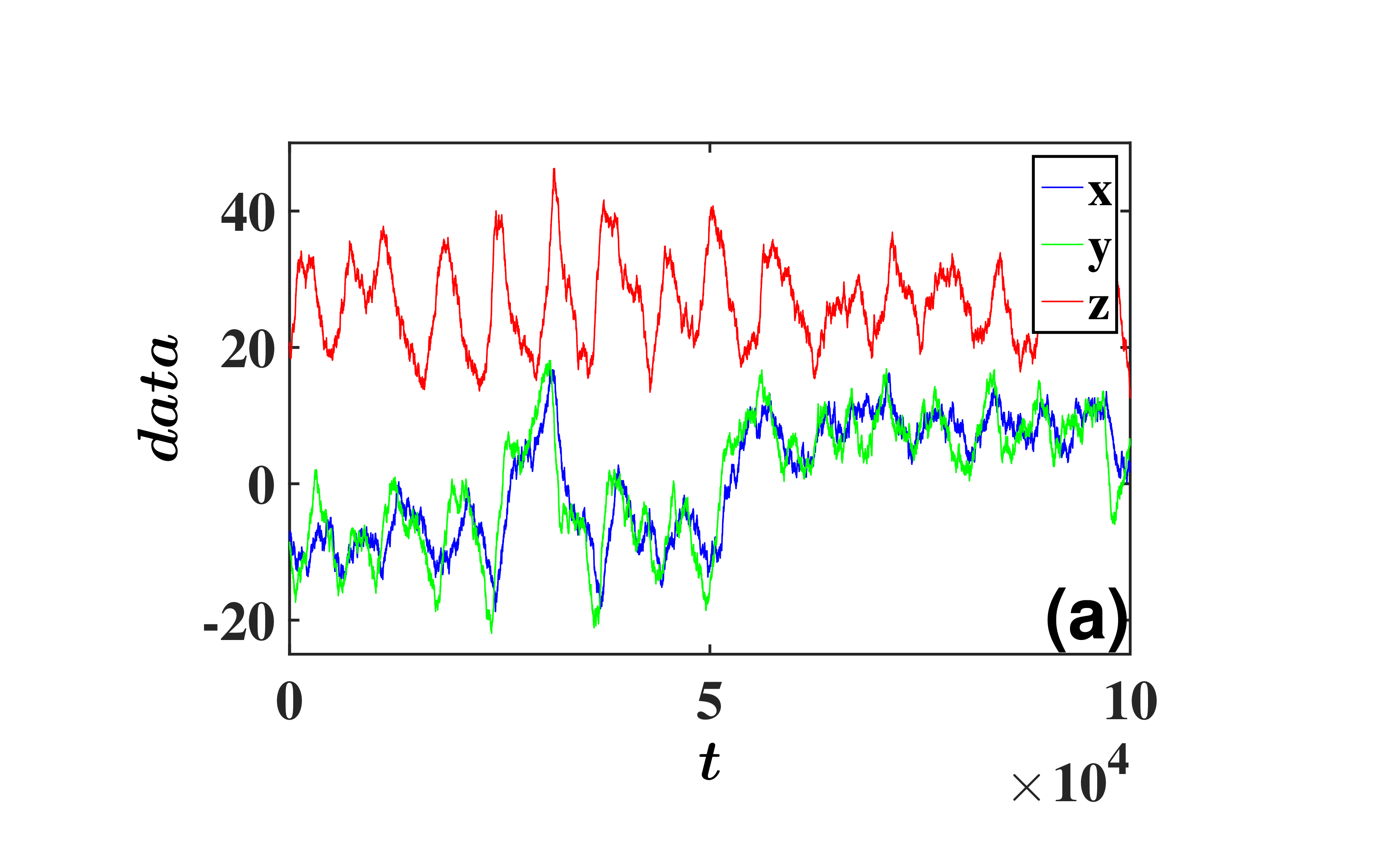}}
    \hfill
  \end{minipage}
  \begin{minipage}[c]{0.48\columnwidth}
    \centering
    \subfigure{\includegraphics[angle=-90,width=0.96\textwidth,totalheight=0.5\textwidth]{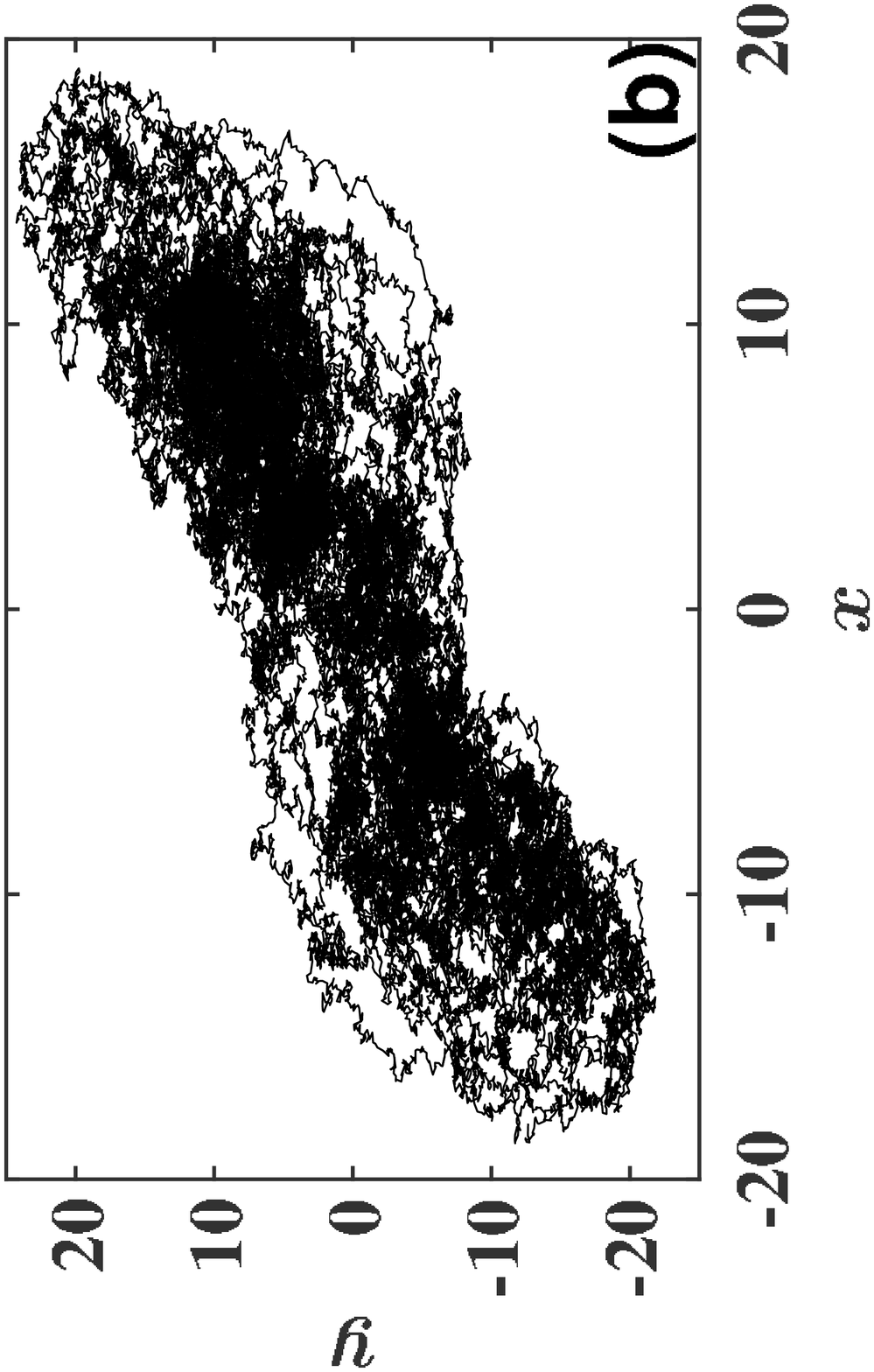}}
  \end{minipage}\\[5pt]
  \begin{minipage}[c]{0.48\columnwidth}
    \centering
    \subfigure{\includegraphics[angle=0,width=0.96\textwidth,totalheight=0.5\textwidth]{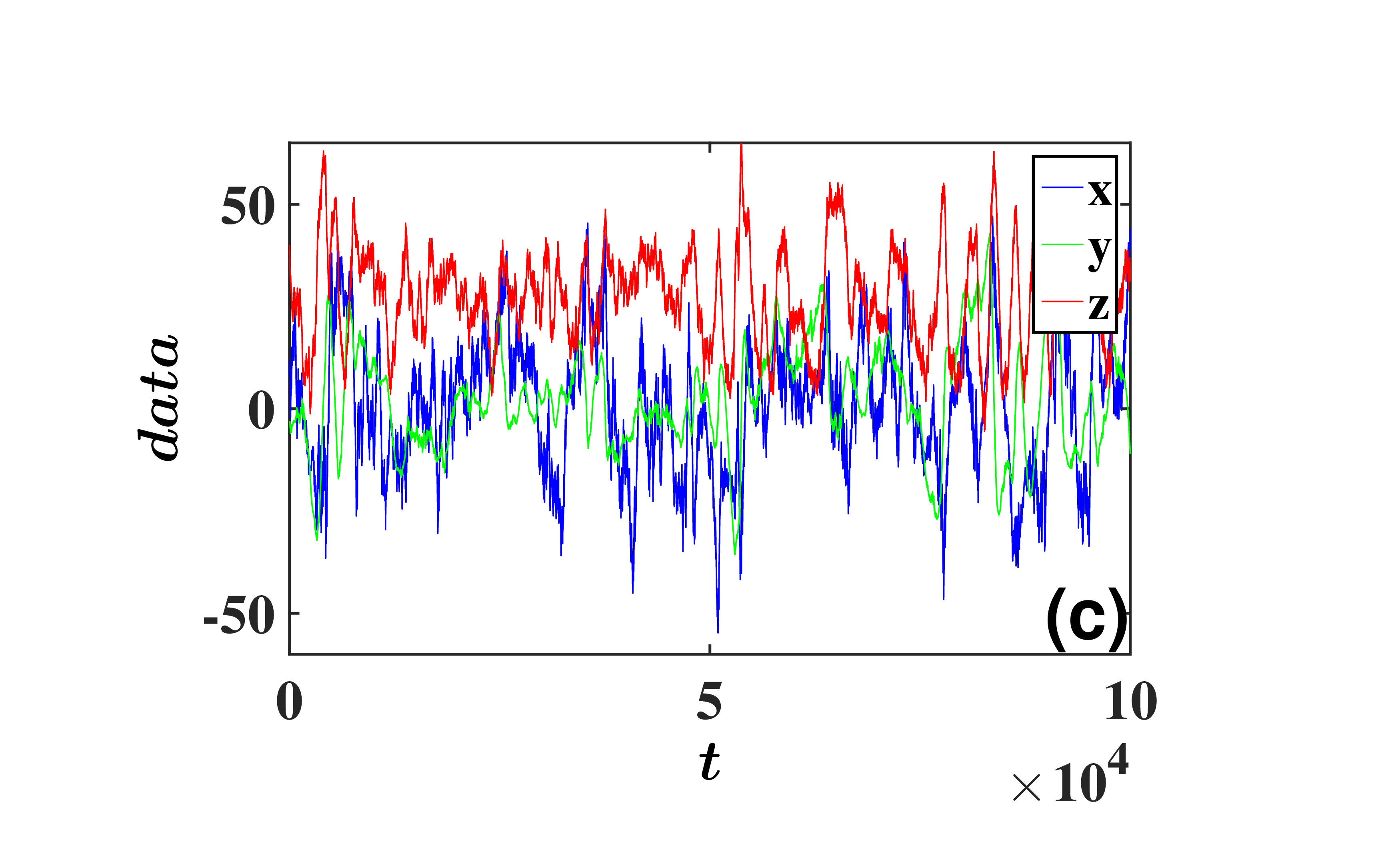}}
    \hfill
  \end{minipage}
  \begin{minipage}[c]{0.48\columnwidth}
    \centering
    \subfigure{\includegraphics[angle=-90,width=0.96\textwidth,totalheight=0.5\textwidth]{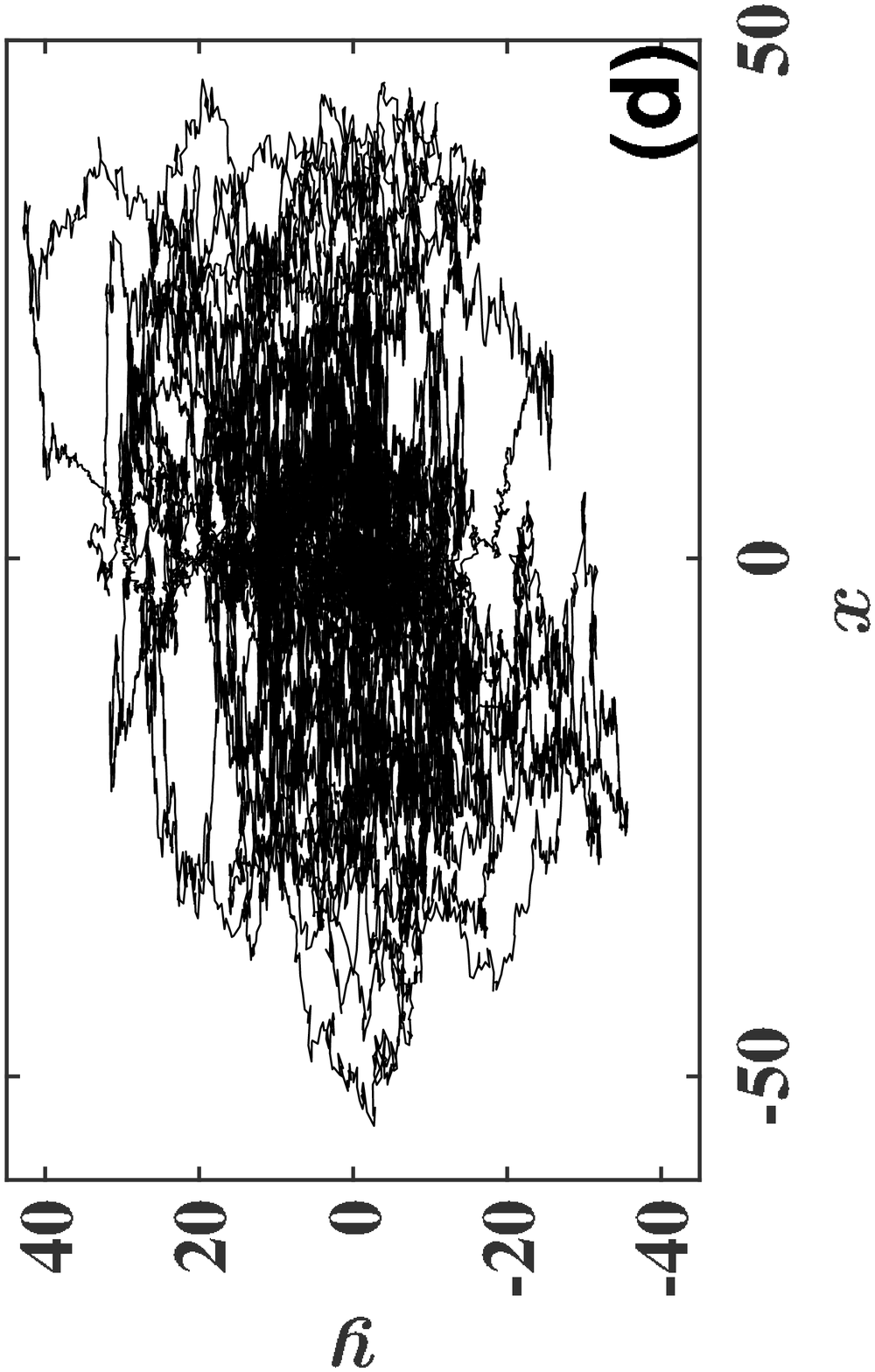}}
  \end{minipage}
  \caption{Time sequences and trajectories of noise-driven Lorentz system.
  Parameters are taken as $\sigma=10$, $\rho=28$, $\beta=8/3$.
  (\textbf{a})(\textbf{b}) Additive noises $Q_{ij}=50\delta_{ij}$ are applied.
  (\textbf{c})(\textbf{d}) Multiplicative noises $a_1=10$, $a_2=-10$, $b_1=10$, $b_2=-5$, $c_1=20$, $c_2=1$ in Eq.(\ref{QMLeq}) are applied.
  Data are chaotic and strongly random.
  Trajectory in (d) is considerably different from that of (b) for different noise factors,
  though Lorentz dynamic field is not changed in all (a)-(d).
  Any deterministic solutions,
  ignoring noise effects are not successful.
  }
  \label{fig1}
\end{figure}

  \section{Inference of noise-driven networks}

  We first consider a three-node nonlinear network, the Lorenz system, one of the most famous models in chaos study \cite{N19}
  \begin{align}
    \dot{x} &= f_1(x,y,z)+\Gamma_1(t) & f_1 &= \sigma(y-x) \nonumber \\
    \dot{y} &= f_2(x,y,z)+\Gamma_2(t) & f_2 &= x(\rho-z)-y  \label{LorenzEq}\\
    \dot{z} &= f_3(x,y,z)+\Gamma_3(t) & f_3 &= xy-\beta z \nonumber
  \end{align}
  with certain given multiplicative noises
  \begin{equation}
    \bm{Q} = \left(\begin{array}{ccc}
      (a_1+a_2\sqrt{|z|})^2 & 0 & 0 \\
      0 & (b_1+b_2\sqrt{|x|})^2 & 0 \\
      0 & 0 & (c_1+c_2\sqrt{|y|}) ^2
    \end{array}\right) \label{QMLeq}
  \end{equation}
  Though network (\ref{LorenzEq}) looks very simple and low-dimensional,
  the inverse problem has not yet been solved so far
  with available time sequences and trajectories shown in Figs.\ref{fig1} where both noise and nonlinearity essentially effect
  and noises with different multiplicative factors can produce considerably different data sets.
  Now we apply the general algorithms Eqs.(\ref{Outcome1}) to this system.
  For start we should choose bases for field expansion.
  Without particular information,
  power series can be naturally chosen as a candidate basis set.
  Then we assume the following bases with truncation $M$,
  \begin{eqnarray}
    \bm{Y}_i & = & \bm{Z}_i = \bm{Y}, \  i=1,2,3 \nonumber\\
    \bm{Y} & = & (Y_1,Y_2,\cdots,Y_M) \label{LExp} \\
     & = & (1,x,y,z,x^2,xy,xz,y^2,yz,z^2,\cdots) \nonumber
  \end{eqnarray}

  Now we introduce an idea of self-consistent checking of the truncation.
  At first we take $M_i^0$ bases in succession of Eq.(\ref{LExp}) and compute all elements of corresponding $\hat{\bm{C}_i}$ and $\bm{B}_i$.
  Then we consider some more bases, i.e., $M_i^1>M_i^0$ bases in Eq.(\ref{LExp}) as the truncated basis set,
  and obtain further results of $\{\bm{A}_i(M_i^1)|M_i^1>M_i^0\}$.
  $M_i^0$ is concluded as a suitable truncation if the results satisfy two conditions.
  Condition (i): $\bm{A}_i(M_i^1)\approx\bm{A}_i(M_i^0)$ for all coefficients obtained with $M_i^0$ truncation;
  Condition (ii): $\bm{A}_i(M_i^1)\approx0$ for all coefficients of bases not included by $M_i^0$ truncation.
  We conclude that $M_i^0$ is a proper truncation.
  If any of the above two conditions is not satisfied we should go on to include more bases in (\ref{LExp}),
  $M_i^2>M_i^1$, $M_i^3>M_i^2$ and so on, till the two conditions are satisfied together at $M_i^k$ truncation.
  We then conclude $M_i^{k-1}$ is the suitable truncation.
  Compressive sensing methods are possible to be used for technically fasting the above truncation process \cite{N9-1,CS1,CS2}.
  In this paper we just use the basic method of seeking truncations from low orders to high orders.
  \begin{figure}
  \begin{minipage}[c]{0.48\columnwidth}
    \centering
    \subfigure{\includegraphics[angle=0,width=0.96\textwidth,totalheight=0.5\textwidth]{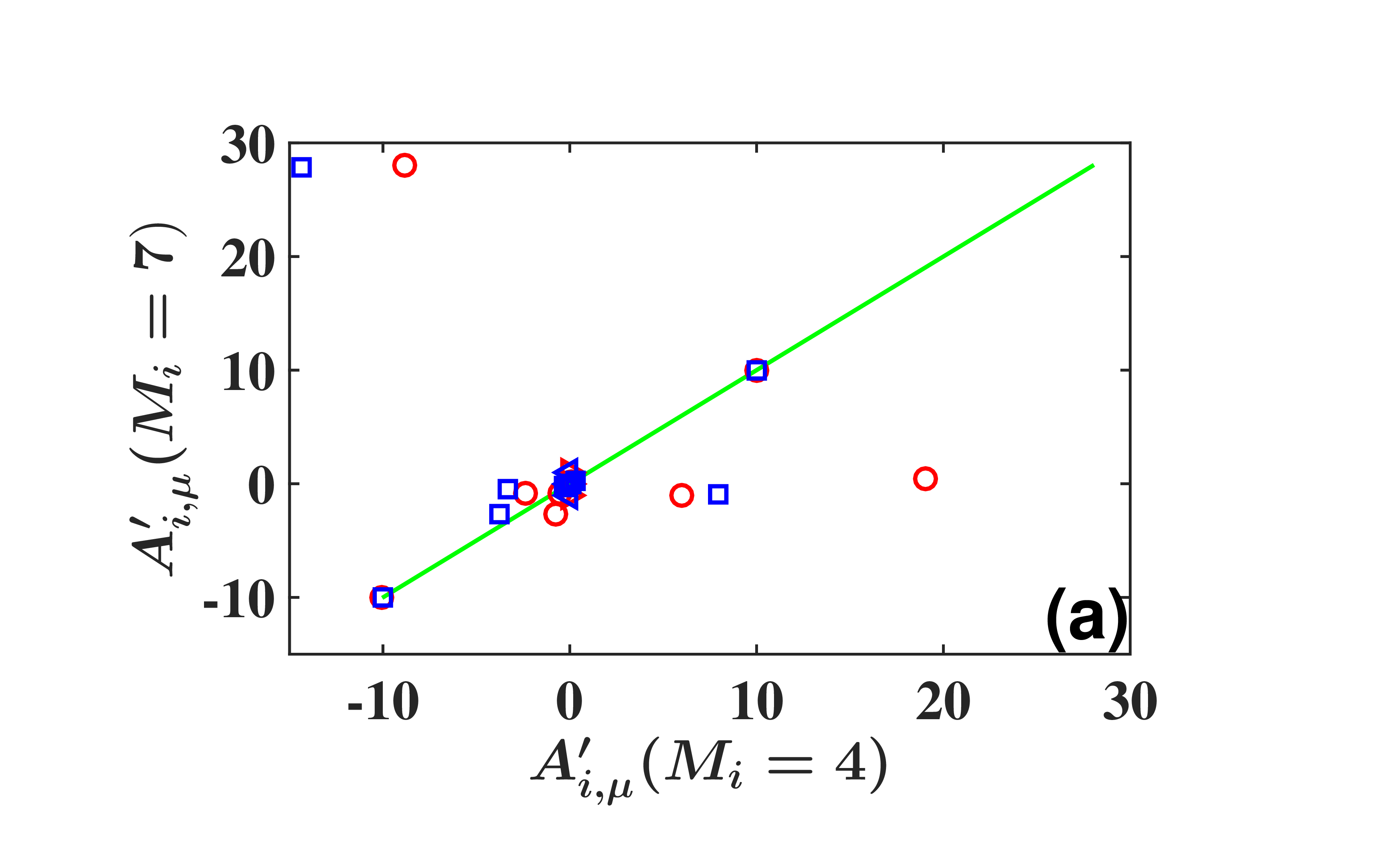}}
    \hfill
  \end{minipage}
  \begin{minipage}[c]{0.48\columnwidth}
    \centering
    \subfigure{\includegraphics[angle=0,width=0.96\textwidth,totalheight=0.5\textwidth]{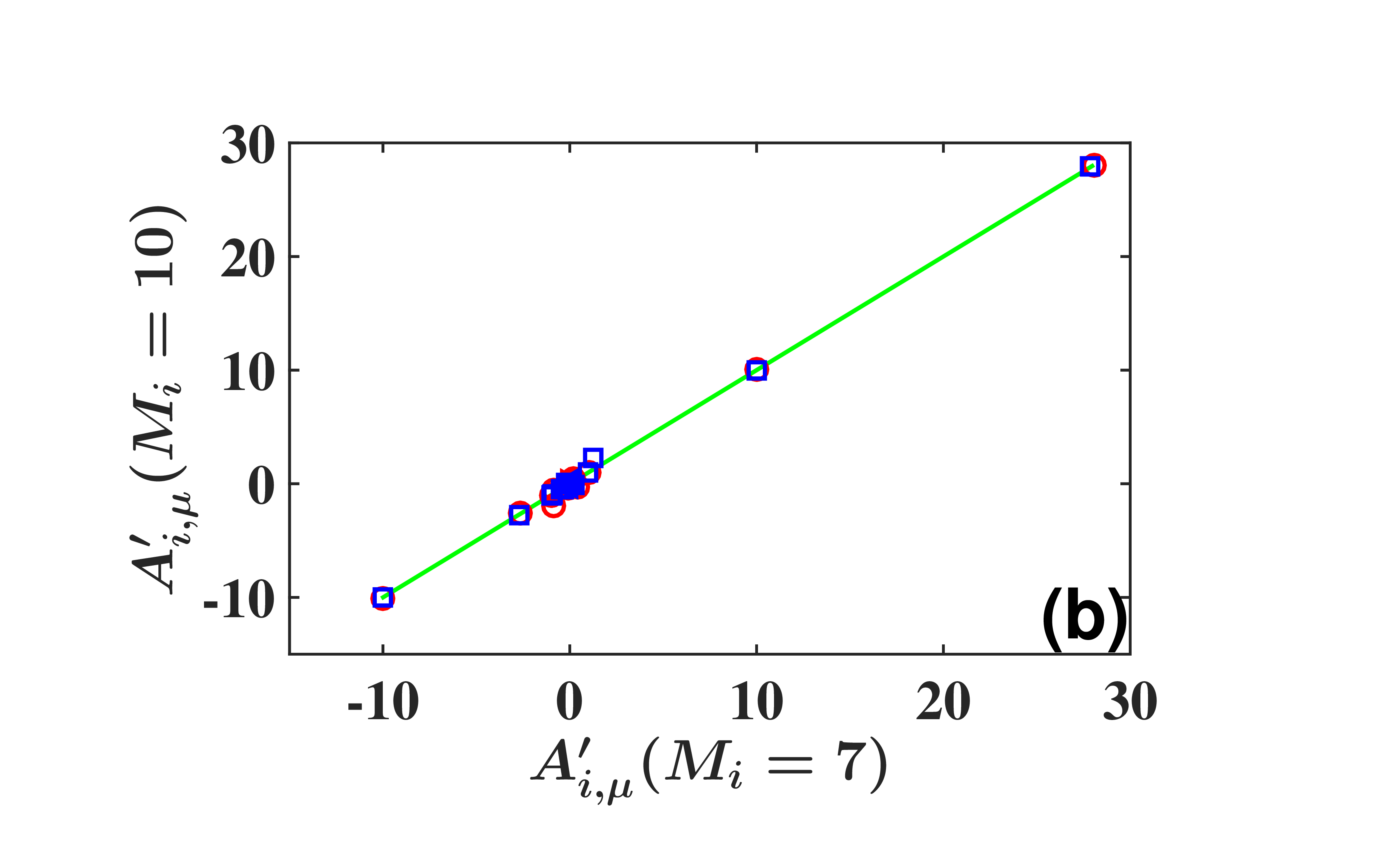}}
  \end{minipage}\\[5pt]
  \begin{minipage}[c]{0.48\columnwidth}
    \centering
    \subfigure{\includegraphics[angle=0,width=0.96\textwidth,totalheight=0.5\textwidth]{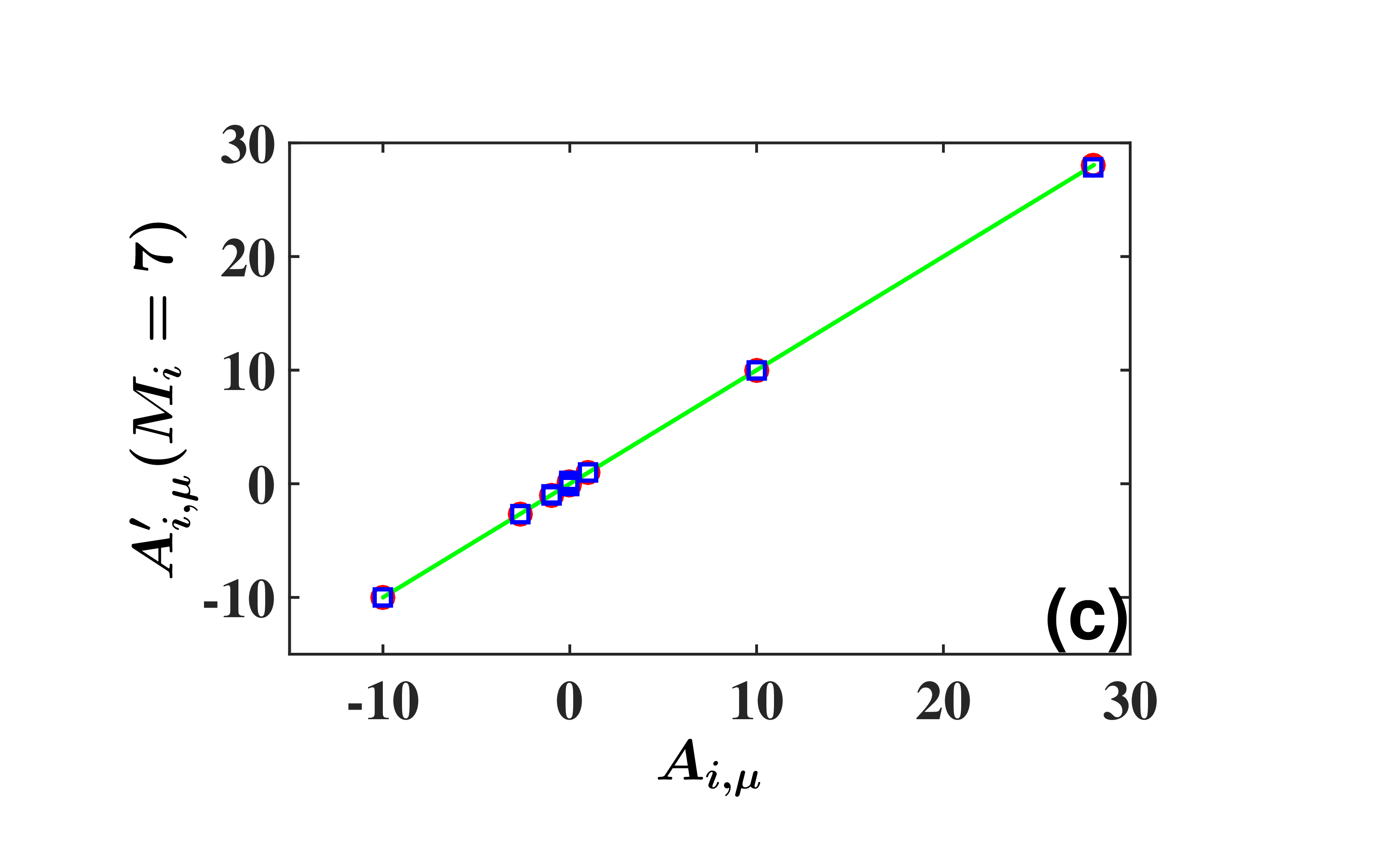}}
    \hfill
  \end{minipage}
  \begin{minipage}[c]{0.48\columnwidth}
    \centering
    \subfigure{\includegraphics[angle=0,width=0.96\textwidth,totalheight=0.5\textwidth]{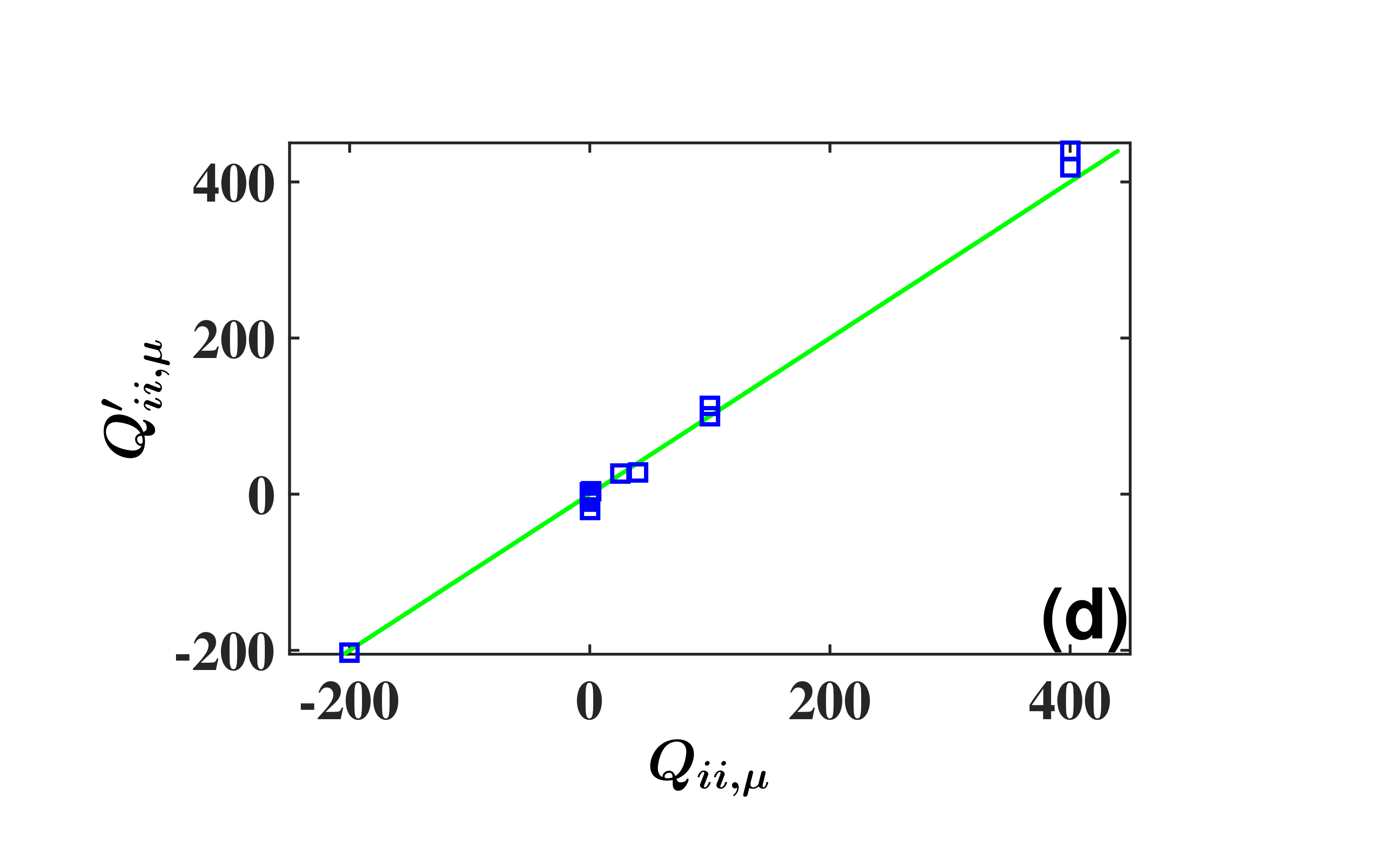}}
  \end{minipage}
  \caption{Inference computations of Eq.(\ref{LorenzEq}).
  Red circles are for additive noises while blue squares for multiplicative ones.
  (\textbf{a}) Depiction results obtained by Eq.(\ref{Outcome1}) with $M_i=4$ (only consider 1 and linear bases) plotted against that with $M_i=7$ (add bases $x^2$, $xy$ and $xz$).
  (\textbf{b}) The same as (a) with $M_i=7$ plotted against that with $M_i=10$ (include all quadratic bases).
  (\textbf{c}) Depiction results obtained by Eq.(\ref{Outcome1}) with $M_i=7$ plotted against the actual coefficients,
  and all plots are around the diagonal line,
  indicating correct depiction at $M_i=7$.
  (\textbf{d}) Depiction of $Q_{ii}$ derived by Eq.(\ref{QEq}) is plotted against actual ones.
  All plots are around the diagonal line,
  indicating the successfulness of the depiction of noise statistics.
  }
  \label{fig2}
\end{figure}

  A self-consistent justification on correct truncation for model Eq.(\ref{LorenzEq}) is illuminated in Fig.\ref{fig2}.
  In Fig.\ref{fig2}(a), it is apparent that $M_1=M_2=M_3=4$ is not a proper truncation,
  since the depiction is considerably different from that of $M_i=7$.
  However, the depiction can well satisfy the above two criterions after $M_i>7$.
  Therefore we can conclude it is enough to truncate the expansion at $M_i=7$ with properly chosen bases in Eq.(\ref{LExp}).
  In Fig.\ref{fig1}(c) we compare the depiction results for $M_i=7$ with the actual $A_{i,\mu}$,
  and the nonlinear and interactive structures of Eq.(\ref{LorenzEq}) are recovered with very high precision.

  For depicting noise statistics we construct expansion and correlators in the similar ways as $Y_i=Z_i$ in Eq.(\ref{LExp}).
  By applying Eqs.(\ref{QEq})(\ref{QEdef}) the coefficients of multiplicative factors of noises $G_{ij,k}$ can be computed.
  The results are shown in Fig.\ref{fig2}(d).
  It is sprising that with the randomly behaved data of Fig.\ref{fig1} only we can correctly explore not only the nonlinear fields and interaction structures,
  but also the detailed multiplicative behaviors of noises.
  \begin{figure}
  \begin{minipage}[c]{0.48\columnwidth}
    \centering
    \subfigure{\includegraphics[angle=-90,width=0.96\textwidth,totalheight=0.5\textwidth]{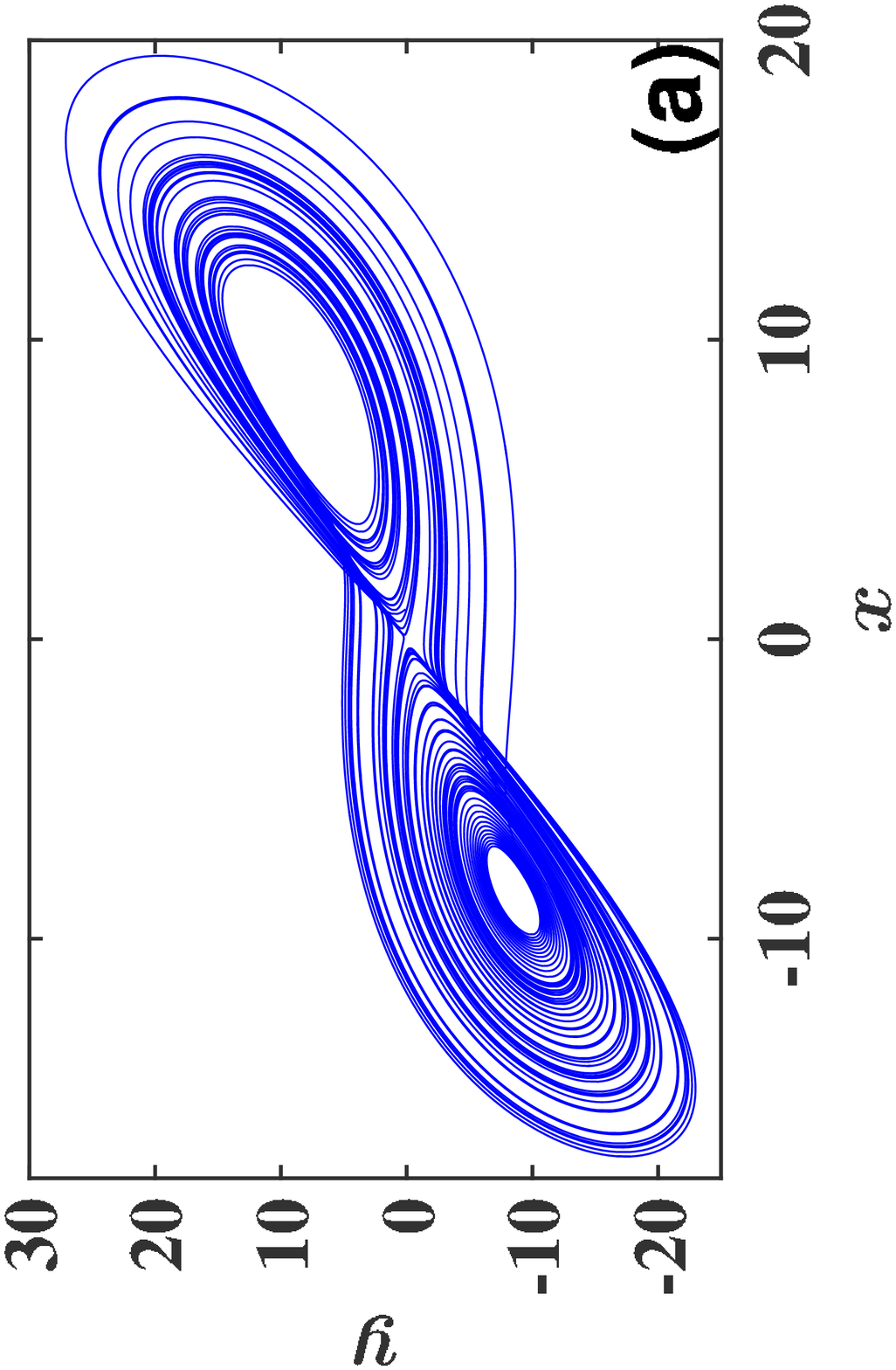}}
    \hfill
  \end{minipage}
  \begin{minipage}[c]{0.48\columnwidth}
    \centering
    \subfigure{\includegraphics[angle=-90,width=0.96\textwidth,totalheight=0.5\textwidth]{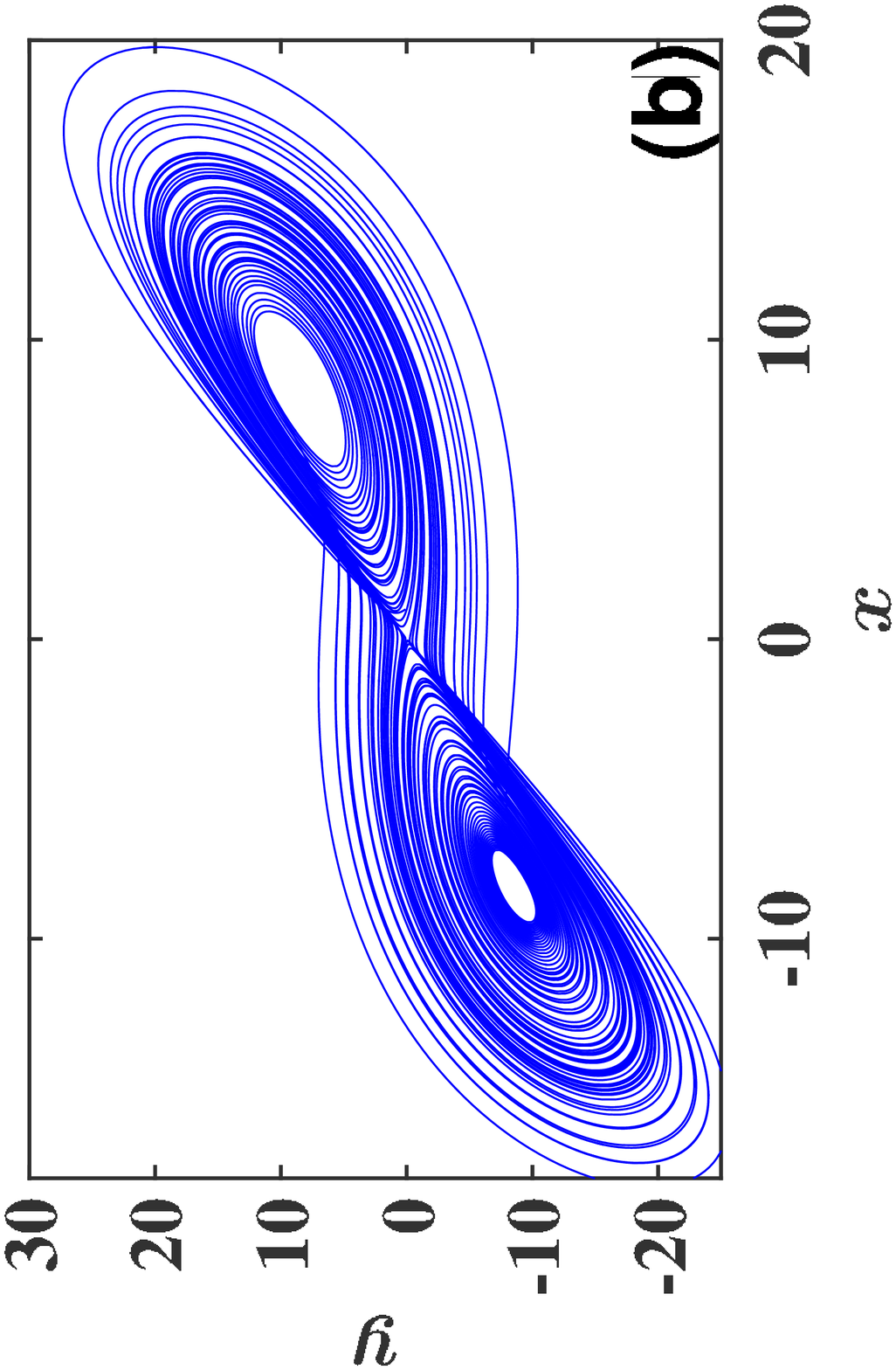}}
    \hfill
  \end{minipage}\\[5pt]
  \begin{minipage}[c]{0.48\columnwidth}
    \centering
    \subfigure{\includegraphics[angle=-90,width=0.96\textwidth,totalheight=0.5\textwidth]{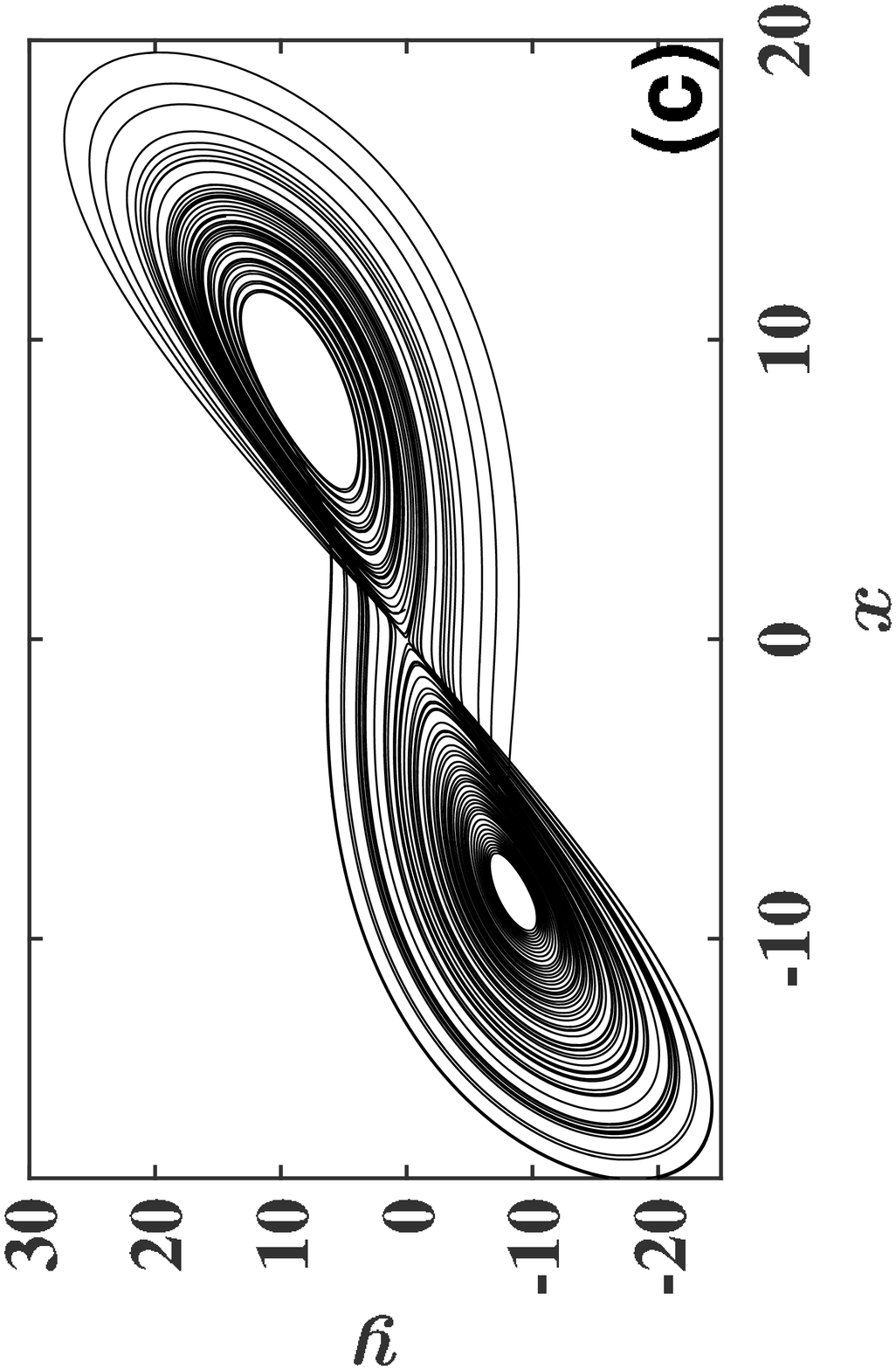}}
  \end{minipage}
  \begin{minipage}[c]{0.48\columnwidth}
    \centering
    \subfigure{\includegraphics[angle=-90,width=0.96\textwidth,totalheight=0.5\textwidth]{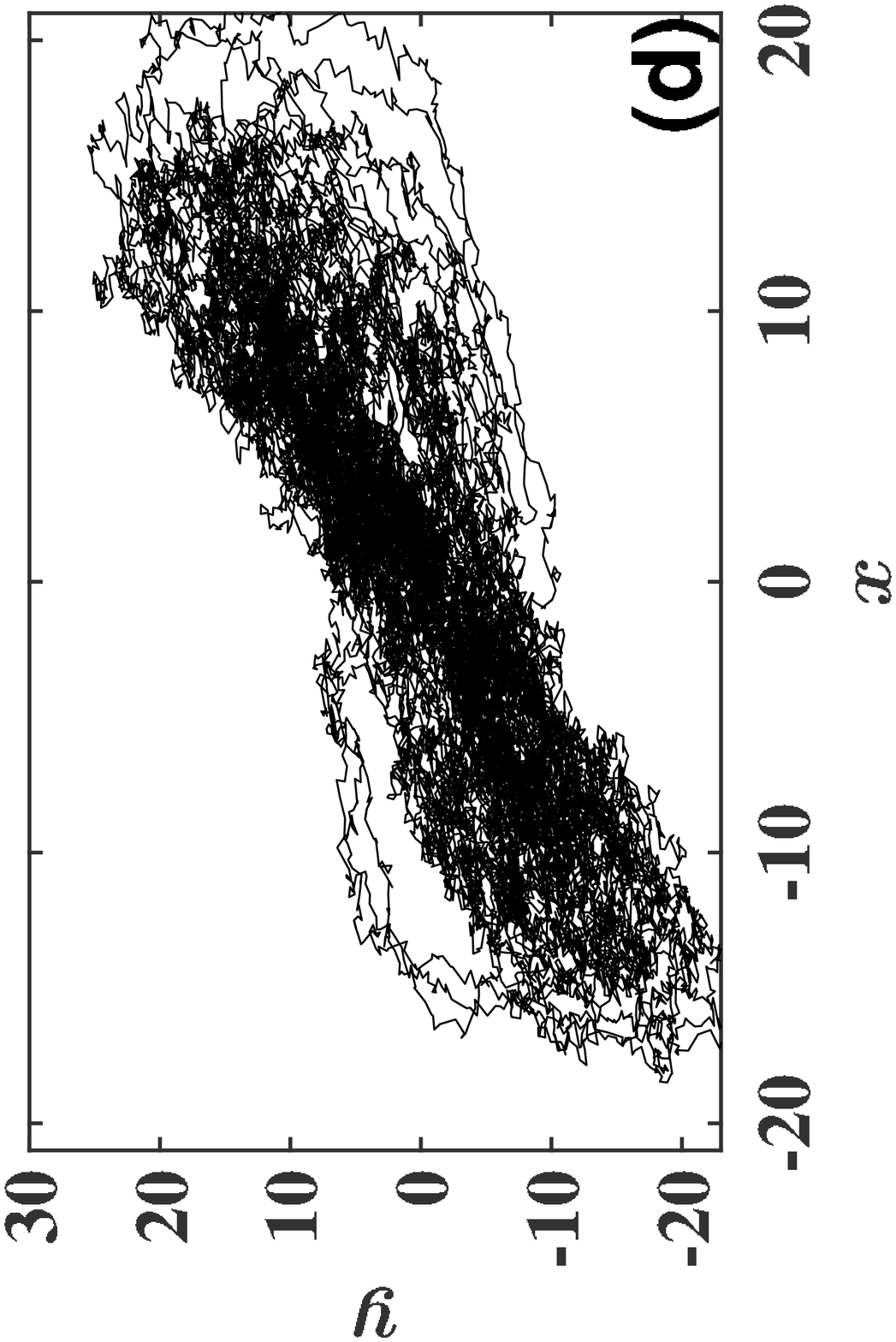}}
    \hfill
  \end{minipage}\\[5pt]
  \begin{minipage}[c]{0.48\columnwidth}
    \centering
    \subfigure{\includegraphics[angle=-90,width=0.96\textwidth,totalheight=0.5\textwidth]{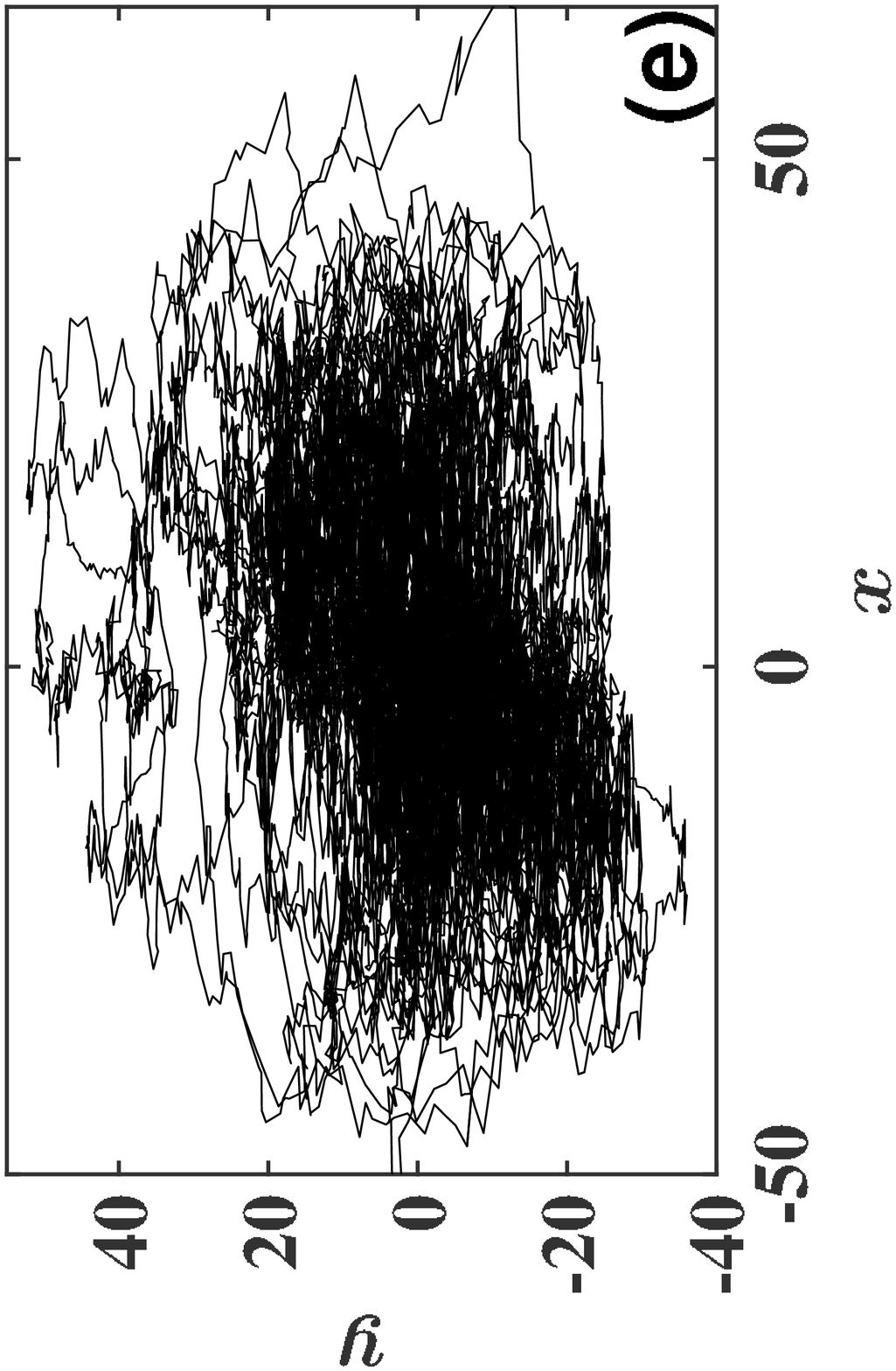}}
  \end{minipage}
  \caption{
  (\textbf{a})(\textbf{b})(\textbf{c}) Comparison of the reproduced trajectories of the system depicted by our HOCC method
   ((a) for additive noises, (b) for multiplicative ones) with that of the original Lorentz system with all noise lifted ((c)).
  Agreement is striking since only the stongly noisy data of Fig.\ref{fig1} are used for dynamics reconstruction,
  and the data in Fig.\ref{fig1}(a)(b) are so different form those in Fig.\ref{fig1}(c)(d).
  (\textbf{d})(\textbf{e}) Reproduced trajectories by models depicted by HOCC with corresponding additive ((d)) and multiplicative ((e)) noises.
  The data in Figs.\ref{fig1} are generated convincingly,
   justifying that the HOCC method is effective for depicting both network structures and noise statistics.
  }
  \label{fig3}
\end{figure}

  We can also use the results of depiction to reconstruct the Lorentz network,
  simulate the depicted dterministic equations ($\bm{f}$ only, without noise) and compare the trajectories (Fig.\ref{fig3}(a)(b)) with the original one with noise lifted (Fig.\ref{fig3}(c)).
  It is found that both reconstructed patterns match the original one well,
  though the two measured trajectories of Figs.\ref{fig1}(a)(b) and (c)(d) differ from each other considerably due to different multiplicative factors of noises.
  This coincidence justifies well the effectiveness of HOCC method.
  In Figs.\ref{fig3}(d)(e) we produce variable data by simulating
  the reconstructed networks with derived noises of the computed multiplicative factors.
  The features of the noisy data are reconstructed perfectly as well.

  Next we infer a network with much higher dimension and more complicated nonlinearities
  \begin{equation}
    \dot{x}_i(t) = \Phi_i(x_i)+\sum_{j=1}^NW_{ij}x_j+\Gamma_i(t),\ i,j=1,2,\cdots,N \label{NEq}
  \end{equation}
Eq.(\ref{NEq}) is very common in practical systems where local dynamics of each node is strongly nonlinear,
and dynamic structures are diverse for different nodes.
On the other hand interactions between nodes, the external facts for each node, are approximated to be linear.
Expanding $\Phi_i(x_i)$ by power series to a power $x_i^{m_i}$
\begin{equation}
  \Phi_i(x_i)=\sum_{\mu=1}^{m_i+1}\alpha_{i,\mu} x_i^{\mu-1} \nonumber
\end{equation}
we have
\begin{eqnarray}
  f_i(\bm{x}) & = & \sum_{\mu=1}^{M_i}A_{i,\mu}Y_{i,\mu} \nonumber \\
   & = & \sum_{\mu=1}^{N}A_{i,\mu}x_\mu+ A_{i,N+1}+\sum_{\mu=N+2}^{M_i}A_{i,\mu}x_i^{\mu-N}
\end{eqnarray}
Then we can represent both unknown nonlinear structure and linear interaction topology unified as
\begin{eqnarray}
  \bm{A}_i & = & (A_{i,1},A_{i,2},\cdots,A_{i,i-1},A_{i,i},A_{i,i+1},\cdots, \nonumber \\
   &  & A_{i,N},A_{i,N+1},A_{i,N+2},\cdots,A_{i,M_i}) \label{NDef1}
\end{eqnarray}
and define basis vector as
\begin{eqnarray}
  \bm{Y}_i(t) & = & (Y_1(t),Y_2(t),\cdots,Y_{M_i}(t)) \label{NDef2} \\
  & = & (x_1,x_2,\cdots,x_N,1,x_i^2,\cdots,x_i^{m_i}) \nonumber
\end{eqnarray}
Choosing correlator vector $\bm{Z_i}$ as $\bm{Z_i}=\bm{Y_i}$, we can specify vector $\bm{B}_i$ and matrix $\hat{\bm{C}}_i$ as
\begin{eqnarray}
  B_{i,\mu} & = & <\dot{x}_i(t)Y_{i,\mu}(t-\tau)> \label{NDef3} \\
  \hat{C}_{i,\mu\nu} & = & <Y_{i,\mu}(t)Y_{i,\nu}(t-\tau)> \nonumber
\end{eqnarray}
Inserting Eq.(\ref{NDef3}) into Eq.(\ref{Outcome1}) we can solve all the unknown elements, including all the nonlinear and linear facts.
\begin{figure}
  \begin{minipage}[c]{0.48\columnwidth}
    \centering
    \subfigure{\includegraphics[angle=0,width=0.9\textwidth,totalheight=0.5\textwidth]{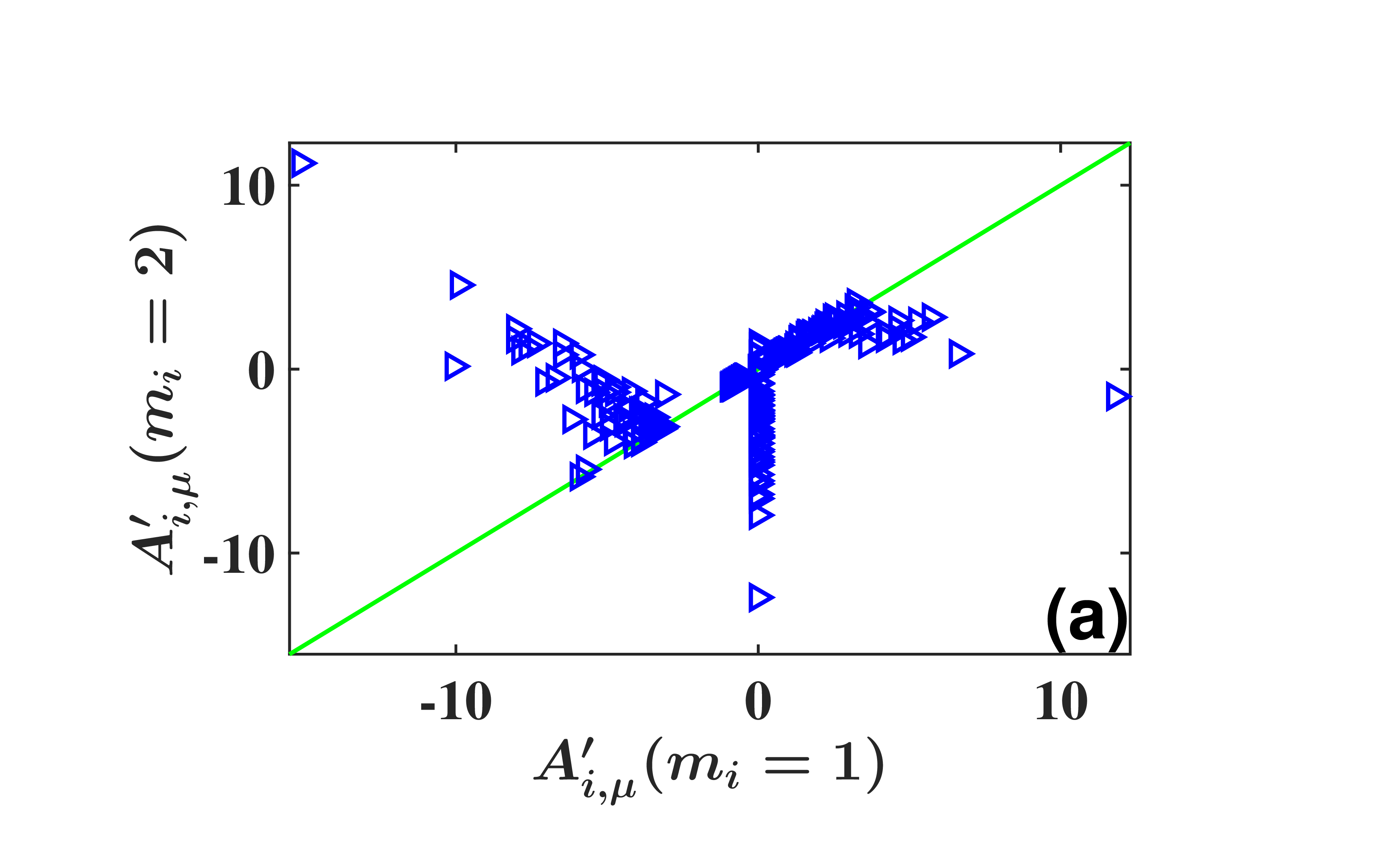}}
    \hfill
  \end{minipage}
  \begin{minipage}[c]{0.48\columnwidth}
    \centering
    \subfigure{\includegraphics[angle=0,width=0.9\textwidth,totalheight=0.5\textwidth]{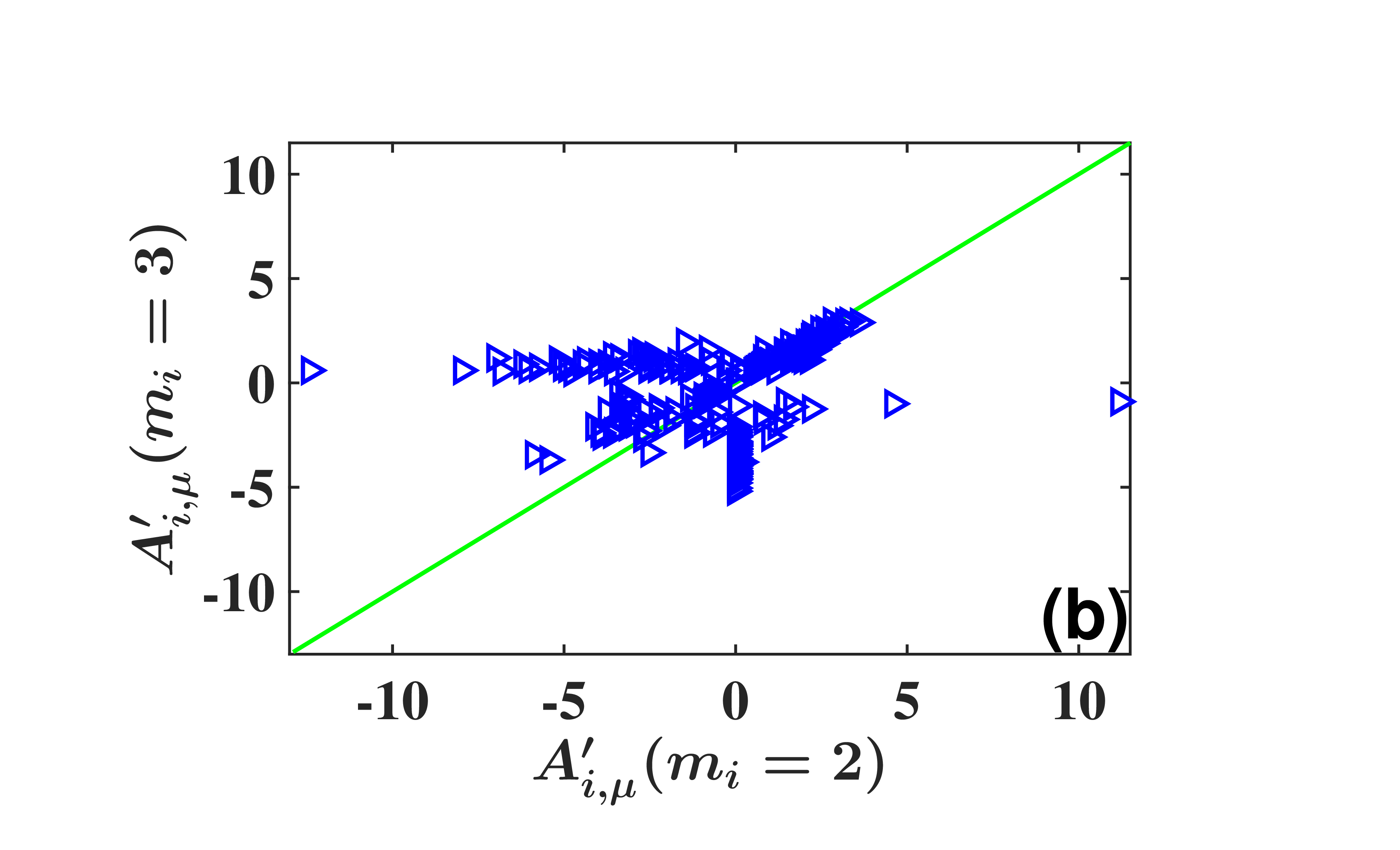}}
  \end{minipage}\\[5pt]
  \begin{minipage}[c]{0.48\columnwidth}
    \centering
    \subfigure{\includegraphics[angle=0,width=0.9\textwidth,totalheight=0.5\textwidth]{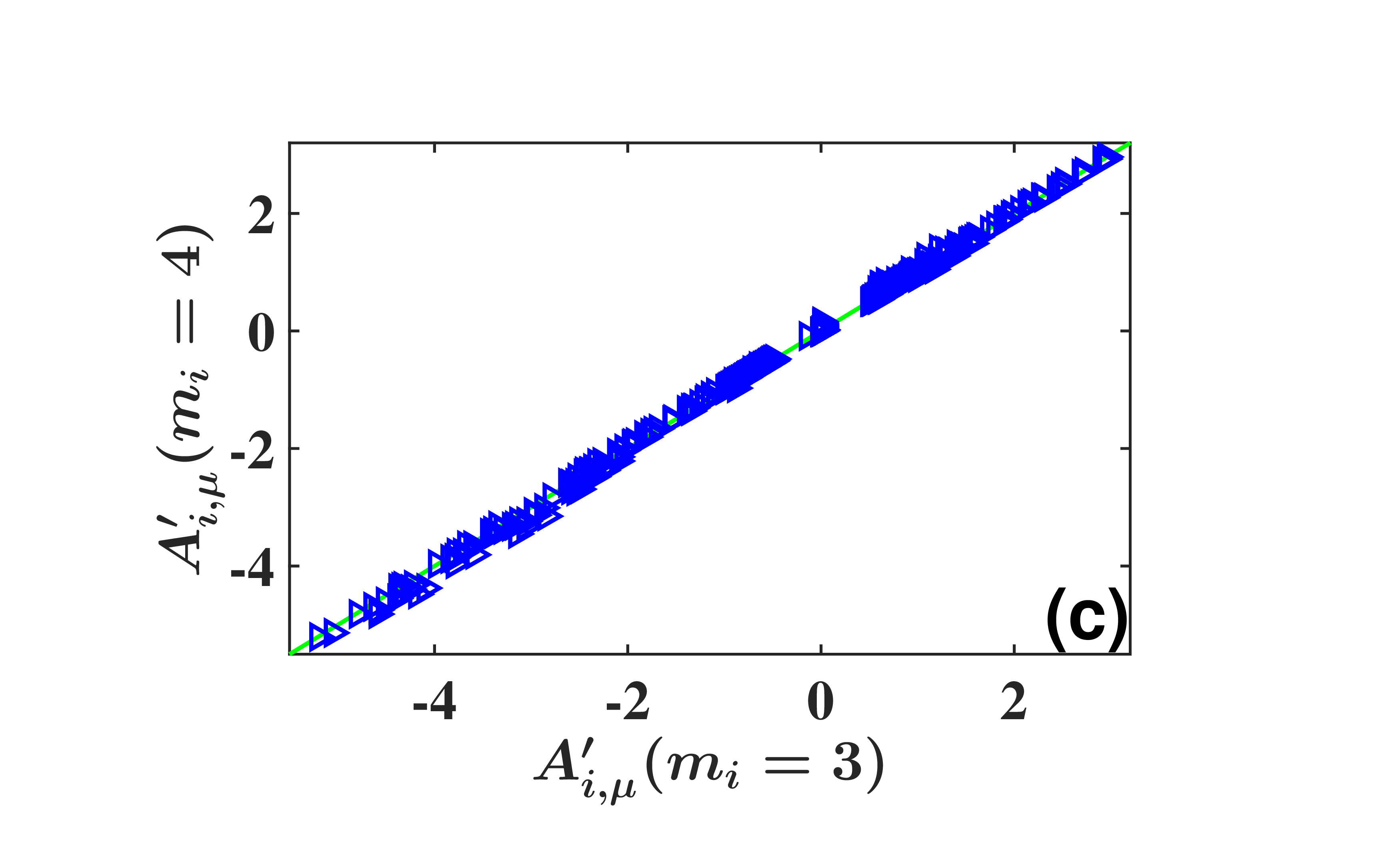}}
  \end{minipage}
  \begin{minipage}[c]{0.48\columnwidth}
    \centering
    \subfigure{\includegraphics[angle=0,width=0.9\textwidth,totalheight=0.5\textwidth]{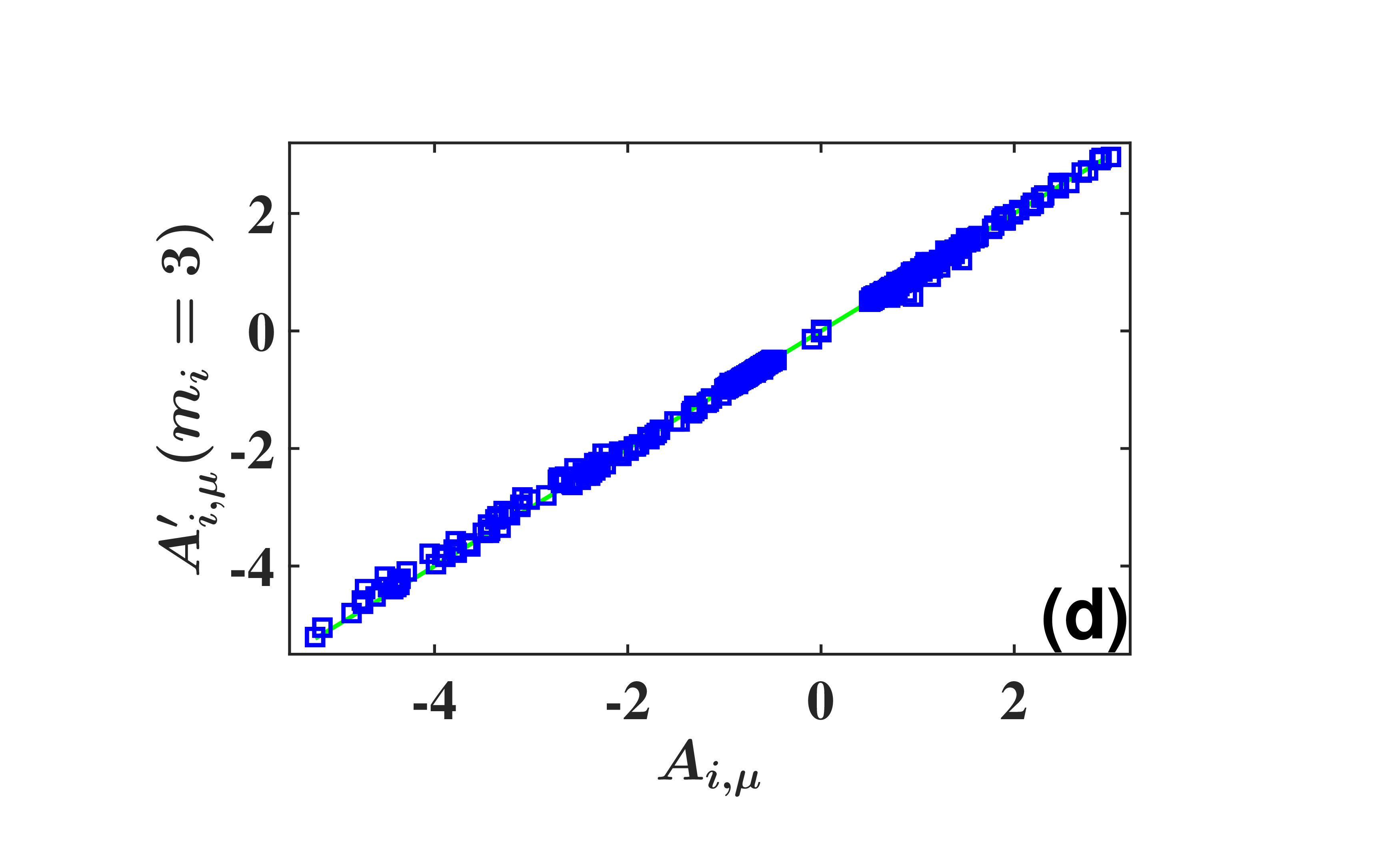}}
  \end{minipage}
  \caption{Depiction of nonlinear structures and linking topology of network in Eq.(\ref{NEq}).
  Parameters are taken with uniform distributions in the intervals $a_i\in(-1,1)$, $b_i\in(2,5)$ and $c_i\in(0,2)$ and
  additive noises $Q_{ij}\in\delta_{ij}(0.5,1)$.
  The interaction intensities are $W_{ij}$ ($i,j=1,2,\cdots,N$) $\in(0.5,1)$ with 10\% probability for positive ones;
  $\in(-1,-0.5)$ also with 10\% for negative ones;
  and $W_{ij}=0$ otherwise.
  Power bases are organized from low-order to higher-order ones.
  (\textbf{a}) Results of HOCC for $m_i=2$ plotted vs. those for $m_i=1$. The dots are considerably away from the diagonal line, representing incorrect truncation at $m_i=1$.
  (\textbf{b}) The same as (a) with results of $m_i=3$ plotted with those of $m_i=2$.
  $m_i=2$ truncation is not suitable either.
  (\textbf{c}) Results for $m_i=4$ plotted v.s. those for $m_i=3$.
  Both results coincide with each other fairly well, that self-consistently confirms the correctness of the HOCC method for sufficiently large $m_i\ge3$.
  (\textbf{d}) The results of HOCC for $m_i=3$ plotted vs. the actual values of $A_{i\mu}$. With suitable nonlinearity considered,
  all nonlinear and interacting structures are depicted correctly with certain fluctuations.  }
  \label{fig4}
\end{figure}

In Fig.\ref{fig4} we investigate a particular case with local dynamics
\begin{equation}
  \Phi_i(x_i) = a_ix_i-b_ix_i^3+c_ie^{-x_i} \label{NLocalDyn}
\end{equation}
with $a_i$,$b_i$,$c_i$ uniformly distributed as $a_i\in(-1,1)$,$b_i\in(2,5)$,$c_i\in(0,2)$.
The network has positive mutual interaction $W_{ij}$, $i,j=1,2,\cdots,N$, $\in(0.5,1)$ with 10\% probability, negative one $\in(-1,-0.5)$ also with 10\%,
and $W_{ij}=0$ otherwise.
Simple additive white noises are used for this model:
\begin{equation}
  Q_{ij}\in\delta_{ij}(0.5,1)
\end{equation}
with uniform probability distribution.
We consider different truncations $m_i$'s.
It is obvious that the depictions have large errors for too small $m_i$ in (Figs.\ref{fig4}(a)(b))
and they can be quickly improved by increasing $m_i$, and saturated at sufficiently large $m_i$ (Figs.\ref{fig4}(c)(d), $m_i=3$ is fairly good approximation in our case).
The approach of self-consistent truncation is stable and convincingly confirmed though the actual expansion of Eq.(\ref{NLocalDyn}) has nonzero coefficients for infinitely large $m_i$'s.

\section{Nonlinear network depiction by using different basis vectors}

For inferring linear networks, the basis vectors can be simply chosen as output variables $\bm{Y}=(x_1,x_2,\cdots,x_N)$.
To depict nonlinear networks, the ways to choose basis vectors become diverse.
In Eqs.(\ref{LExp}) and (\ref{NDef2}) power expansions are used for representing nonlinear functions.
Different types of basis vectors can be used, depending on the nature of data and property of nonlinear dynamics.
In many practical cases the inverse computations can be much more simplified by selecting suitable basis vectors.
Let us consider a network of coupled Kuramoto model \cite{N18} which has been extensively studied for describing oscillatory complex systems.
\begin{gather}
  \dot{\theta}_i = w_i+\Phi_i(\theta_i)+\sum_{j\ne i}^N \Psi_{ij}(\theta_j-\theta_i)+\Gamma_i(t) \label{KEq} \\
  \theta_i+2\pi k=\theta_i,\ k=\pm1,\pm2,\cdots,\pm p,\cdots,\ i=1,2,\cdots,N \nonumber
\end{gather}
where $\theta_i$s, $i=1,2,\cdots,N$, represent phase angles of oscillators,
and all $\Phi_i(\theta_i)$, $\Psi_{ij}(\phi)$ are unknown nonlinear functions with topology
\begin{eqnarray}
  \Phi_i(\theta_i+2\pi k) & = & \Phi_i(\theta_i) \label{KDef1} \\
  \Psi_{ij}(\phi+2\pi k) & = & \Psi_{ij}(\phi) \nonumber
\end{eqnarray}
It is not convenient to approximate functions Eq.(\ref{KDef1}) by power expansions while we can conveniently do it by using Fourier basis vectors.
Expanding $\Phi_i$, $\Psi_{ij}$ as
\begin{eqnarray}
  \Phi_i(\theta_i) & = & \sum_{k=1}^{m_i}[\alpha_{i,k}\sin(k\theta_i)+\alpha_{i,k}^\prime \cos(k\theta_i)] \label{KExp} \\
  \Psi_{ij}(\theta_j-\theta_i) & = & \sum_{k=1}^{m_i}[\beta_{ij,k}\sin(k(\theta_j-\theta_i))+ \nonumber \\
  &  &\beta_{ij,k}^\prime\cos(k(\theta_j-\theta_i))] \nonumber
\end{eqnarray}
we can then define basis vector as
\begin{subequations}
\label{KDef23}
\begin{eqnarray}
  \bm{Y}^T_i & = & (1,\sin(\theta_1-\theta_i),\cos(\theta_1-\theta_i),\sin(2(\theta_1-\theta_i)), \nonumber \\
  & & \cos(2(\theta_1-\theta_i)), \cdots, \sin(m_i(\theta_1-\theta_i)), \nonumber \\
  & & \cos(m_i(\theta_1-\theta_i)), \sin(\theta_2-\theta_i), \cos(\theta_2-\theta_i),  \nonumber \\
  & & \cdots, \sin(m_i(\theta_2-\theta_i)), \cos(m_i(\theta_2-\theta_i)), \cdots,  \nonumber \\
  & & \sin(\theta_{i-1}-\theta_i), \cos(\theta_{i-1}-\theta_i), \cdots, \nonumber \\
  & & \sin(\theta_i),\cos(\theta_i), \cdots, \sin(m_i\theta_i), \cos(m_i\theta_i), \nonumber \\
  & & \sin(\theta_{i+1}-\theta_i),\cdots, \sin(\theta_N-\theta_i), \cos(\theta_N-\theta_i), \nonumber \\
  & & \cdots, \sin(m_i(\theta_N-\theta_i)), \cos(m_i(\theta_N-\theta_i)) ) \label{KDef2}
\end{eqnarray}
and the corresponding unknown coefficient vector as
\begin{equation}
  \bm{A}_i  =  (A_{i,1},A_{i,2},\cdots,A_{i,M_i}),\ M_i=2m_iN+1 \label{KDef3}
\end{equation}
The corellator vector $\bm{Z}_i$ can be simply defined as
\begin{equation}
  \bm{Z}_i(t) = \bm{Y}_i(t) \label{KDef4}
\end{equation}
\end{subequations}
Inserting Eq.(\ref{KDef2}) for $\bm{Y}(t)$ and Eq.(\ref{KDef4}) for $\bm{Z}(t-\tau)$ (0$<\tau\ll$1) into Eq.(\ref{Outcome1}),
we can specify all elements of vector $\bm{B}_i$ and matrix $\hat{\bm{C}}_i$,
and explicitly infer all the nonlinear structures and interaction links of targeted vector $\bm{A}_i$ in Eq.(\ref{KDef3}).
The computational formulism is theoretically closed.
\begin{figure}
  \begin{minipage}[c]{0.48\columnwidth}
    \centering
    \subfigure{\includegraphics[angle=-90,width=0.9\textwidth,totalheight=0.5\textwidth]{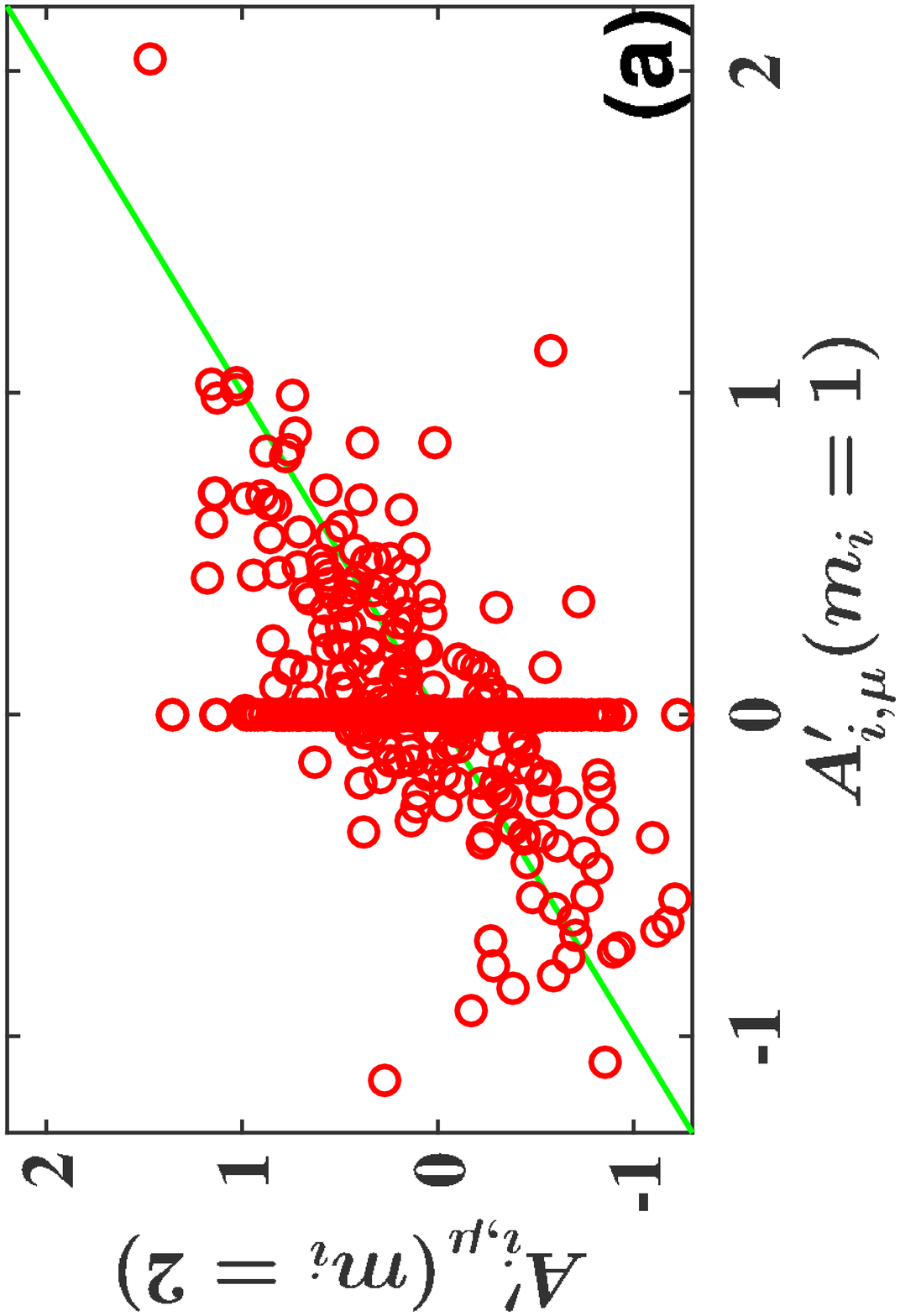}}
    \hfill
  \end{minipage}%
  \begin{minipage}[c]{0.48\columnwidth}
    \centering
    \subfigure{\includegraphics[angle=-90,width=0.9\textwidth,totalheight=0.5\textwidth]{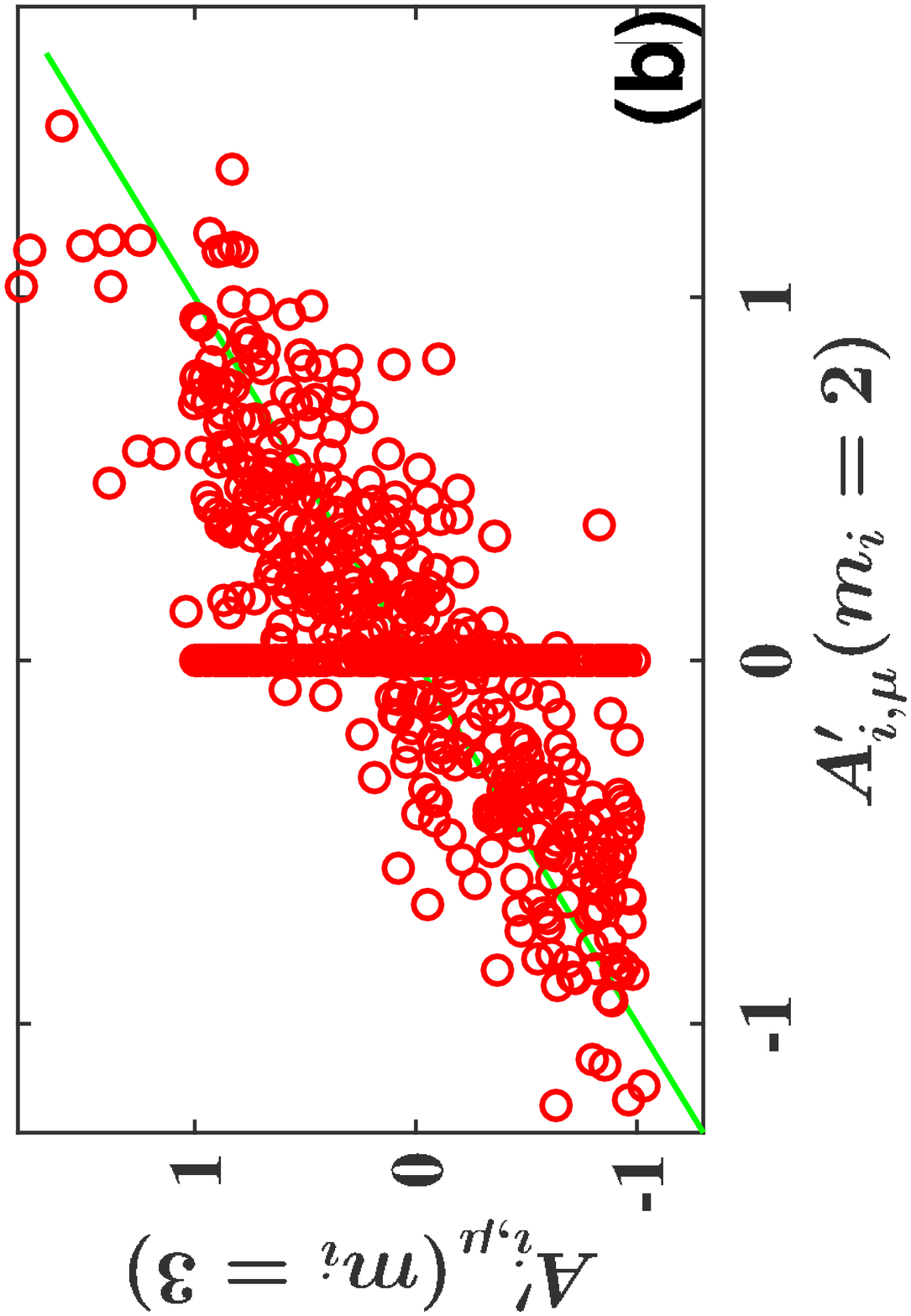}}
  \end{minipage}\\[5pt]
  \begin{minipage}[c]{0.48\columnwidth}
    \centering
    \subfigure{\includegraphics[angle=-90,width=0.9\textwidth,totalheight=0.5\textwidth]{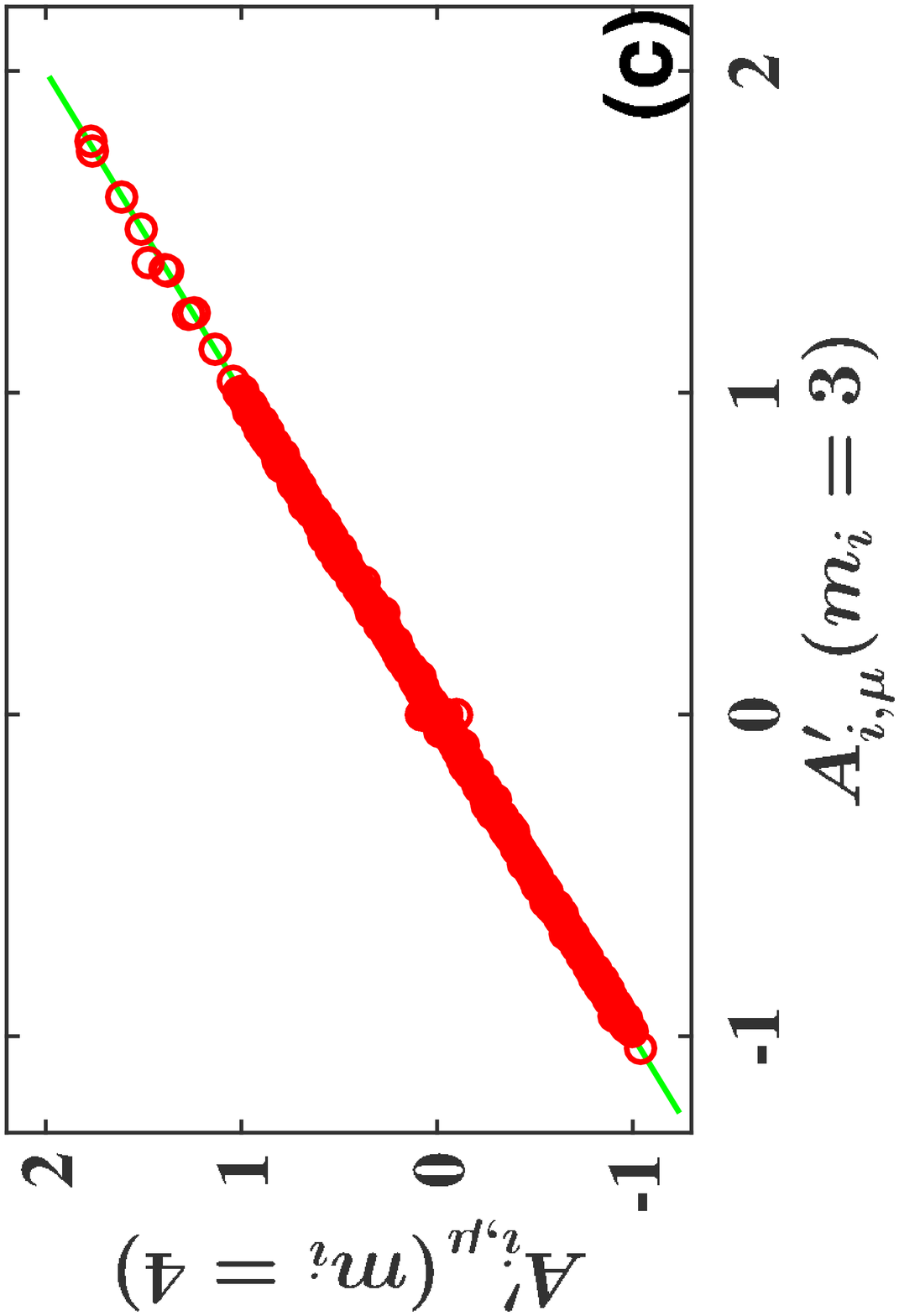}}
  \end{minipage}
  \begin{minipage}[c]{0.48\columnwidth}
    \centering
    \subfigure{\includegraphics[angle=-90,width=0.9\textwidth,totalheight=0.5\textwidth]{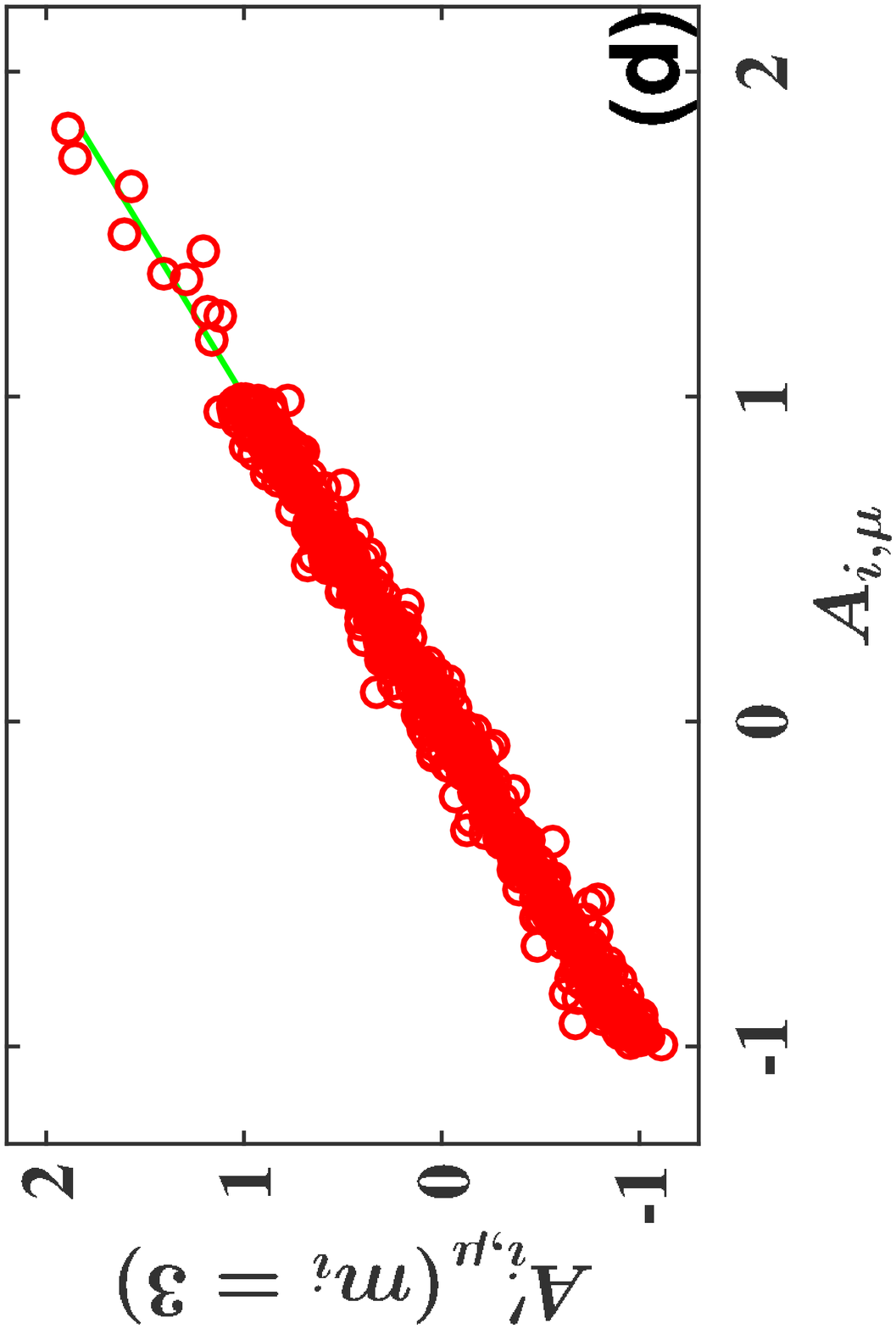}}
  \end{minipage}\\[5pt]
  \begin{minipage}[c]{0.48\columnwidth}
    \centering
    \subfigure{\includegraphics[angle=-90,width=0.9\textwidth,totalheight=0.5\textwidth]{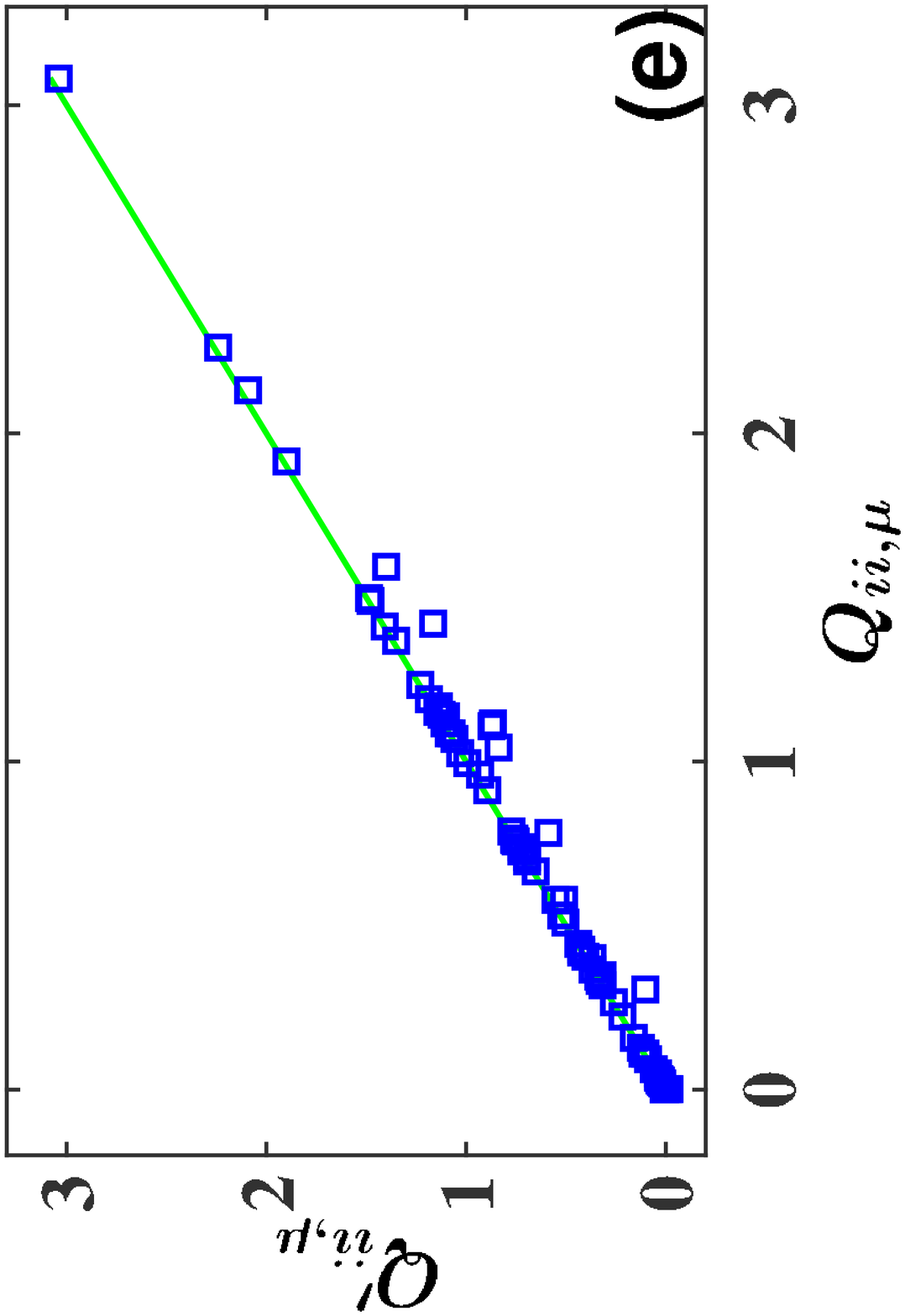}}
  \end{minipage}
  \caption{Inferences of oscillatory network Eq.(\ref{KEq}) by using Fourier basis set.
  $N=10$, and other parameters are uniformly distributed on $A_{i,\mu}$$\in$(-1,1)$(\mu\ge 2)$, $A_{i,1}\in$(1,2) and all the fourier components of $m_i\ge4$ are zero.
  Multiplicative noises Eq.(\ref{QMKeq}) are used with $a_{ij}, b_{ij,k}\in(0,1)$.
  Truncations are made from low-order harmonics to higher-order ones.
  (\textbf{a})(\textbf{b}) The same as Fig.\ref{fig2}(a) with phase dynamics Eq.(\ref{KEq}) considered,
   and the results of $m_i=2$ plotted v.s. $m_i=1$ (a); $m_i=3$ v.s. $m_i=2$ (b). Without high-order harmonics, many plots distribute away from the diagonal lines.
  (\textbf{c}) Results of $m_i=4$ plotted v.s. those of $m_i=3$. All dots are distributed around the diagonal line, indicating self-consistently the suitable truncation at $m_i=3$.
  (\textbf{d}) The same as Fig.\ref{fig2}(c) with Eq.(\ref{KEq}) computed. All depicted results at $m_i=3$ coincide with those of the actual $A_{i,\mu}$.
  (\textbf{e}) The same as Fig.\ref{fig2}(d) with Eq.(\ref{QMKeq}) noises computed.
  The depicted results of multiplicative noises agree rather well with actual ones.
  HOCC method works perfectly for depicting both networks of phase dynamics and noise statistics, and the self-consistent checking works.}
  \label{fig5}
\end{figure}

We take a network of $N=10$ as an example with
\begin{eqnarray}
  \Phi_i(\theta_i) & = & \sum_{k=1}^3[\alpha_{i,k}\sin(k\theta_i)+\alpha_{i,k}^\prime\cos(k\theta_i)] \nonumber \\
  \Psi_{ij}(\theta_j-\theta_i) & = & \sum_{k=1}^3[\beta_{ij,k}\sin(k(\theta_j-\theta_i))+ \nonumber \\
   & & \beta^\prime_{ij,k}\cos(k(\theta_j-\theta_i))], \ i\ne j
\end{eqnarray}
where all $\alpha_{i,k}$, $\alpha_{i,k}^\prime$, $\beta_{ij,k}$, $\beta^\prime_{ij,k}$, $k=1,2,3$ uniformly distribute in the interval $(-1,1)$.
Multiplicative noises $\Gamma_i(t)$s are simply chose as
\begin{equation}
  Q_{ij}(\bm{x}) = [a_{ij}+\sum_{k=1}^2b_{ij,k} (sin(k\theta_i)+cos(k\theta_i))]^2\delta_{ij} \label{QMKeq}
\end{equation}
In Fig.\ref{fig5}(a) the results of inference for $m_i=1$ in which all existing high-order harmonic terms are not considered,
are compared with those for $m_i=2$.
The results are poor.
With $m_i$ increased, the results in Fig.5(b) are improved.
In Fig.\ref{fig5}(c) we compare the depiction results for $m_i=3$ and $m_i=4$, and find that all results of $m_i=3$ are almost not changed in the case of $m_i=4$,
indicating the correctness of the former truncation.
With the changes from Fig.\ref{fig5}(a) to \ref{fig5}(b) to \ref{fig5}(c) we can surely conclude that the results with truncations of $m_i=1$ and $2$ are not correct
while the correctness of $m_i=3$ is justified in a self-consistent way.
In Fig.5(d) the results of inference of $m_i=3$ are compared with actual ones.
With all harmonic terms being taken into account,
we achieve rather precise depiction with certain fluctuations caused by noise and finite measurement frequency and finite data samples.
Moreover, by applying the approaches Eqs.(\ref{QEq})(\ref{QEdef}) we can explore the noise statistical structures and reveal multiplicative factors rather accurately (Fig.\ref{fig5}(e))
\section{Discussion}

In conclusion we have studied the inverse problem of noise-driven nonlinear dynamic networks with measurable data of node variables in networks only.
A high-order correlation computation (HOCC) method is proposed to unifiedly treat nonlinear dynamic structures,
coupling topology and statistics of additive and multiplicative noises in networks.
This method treats inverse problems by jointly considering three facts:
 choosing suitable basis and correlator vectors to  expand nonlinear terms of networks;
 adjusting correlation time difference to treat the noise effects;
 and applying high-order correlations to derive linear matrix equations to explore network structures, topology and noise correlation matrics.
the HOCC algorithm has been theoretically derived,
and its predictions are well confirmed by numerical results.

In biological, social and other crossing fields,
we have extremely rich and huge amount of data available while often understand much less about the underlying mechanisms,
the structures and dynamics producing these data.
Nonlinear dynamics and mutual node interactions often cooperate to yield various functions, and noise often play essential roles in biological and social processes.
Now with the development of the inverse problem research,
it is hopefully expected that we can explore the hidden mechanisms,
find the underlying principles and reveal various key parameters by analyzing measurable data outputed from networks.
These capabilities create a novel and significant perspective for understanding, modulating and controlling realistic network processes. \vfill

\appendix*
\section{Derivation of Eq.(\ref{QEdef})(\ref{QEq})}

  Equations (\ref{QEdef}) and (\ref{QEq}) can be derived as follows:
  \begin{eqnarray}
    S_{ij}(\bm{x},t) & = & \dot{x}_i(t)(x_j(t+\tau)-x_j(t-\tau)) \label{A1} \\
     & = & (f_i(\bm{x},t)+\Gamma_i(t))\int_{t-\tau}^{t+\tau}(f_j(\bm{x},t^\prime)+\Gamma_j(t^\prime))dt^\prime  \nonumber
  \end{eqnarray}
  Its expectation on noise realizations reads
  \begin{eqnarray}
    <S_{ij}(\bm{x},t)> & \approx & 2f_i(\bm{x},t)f_j(\bm{x},t)\tau+2<\Gamma_i(t)>f_j(\bm{x},t)\nonumber \\
    & & +f_i(\bm{x},t)\int_{t-\tau}^{t+\tau}<\Gamma_j(t^\prime)>dt^\prime \nonumber \\
    & & +\int_{t-\tau}^{t+\tau}<\Gamma_i(t)\Gamma_j(t^\prime)>dt^\prime
  \end{eqnarray}
  Since $<\Gamma_i(t)>=<\Gamma_j(t)>$=0 and $\tau\ll 1$ we have
  \begin{eqnarray}
    <S_{ij}(\bm{x},t)> & \approx & \int_{t-\tau}^{t+\tau}Q_{ij}(\bm{x}(t))\delta(t-t^\prime)dt^\prime \nonumber \\
     & = & Q_{ij}(\bm{x}(t)) \label{A2}
  \end{eqnarray}
  considering the expansion of $Q_{ij}(\bm{x})$ on $\bm{q}_{ij}(\bm{x})$ bases
  \begin{equation}
    Q_{ij}(\bm{x}) = \sum_{\mu=1}^{M_{ij}}D_{ij,\mu}q_{ij,\mu}(\bm{x}) \label{A3}
  \end{equation}
  and multiplying the two sides of Eq.(\ref{A2}) by basis $q^\prime_{ij,\mu}(\bm{x}(t-\tau))$ and making time average, we arrive at
  \begin{eqnarray}
    \Delta\bm{B}_{ij} & = & \bm{D}_{ij}\hat{\bm{G}}_{ij} \label{A4}
  \end{eqnarray}
  with $\Delta \bm{B}_{ij}$ being vector having elements
  \begin{eqnarray}
    \Delta B_{ij,\nu} & = & <\dot{x}_i(t)(x_j(t+\tau)-x_j(t-\tau))q_{ij,\nu}^\prime(\bm{x}(t-\tau))> \nonumber  \\
    & = & \frac{1}{L-2p}\sum_{k=p+1}^{L-p}\dot{x}_i(t_k)(x_j(t_{k+p})-x_j(t_{k-p}))\cdot \nonumber \\
    & & q^\prime_{ij,\nu}(x(t_{k-p})) \label{A5} \\
     & & \nu = 1,2,\cdots,M_{ij} \nonumber
  \end{eqnarray}
  and $\hat{\bm{G}}_{ij}$ being matrix with elements
  \begin{equation}
    G_{ij,\mu\nu} = \frac{1}{L-p}\sum_{k=1+p}^L q_{ij,\mu}(\bm{x}(t_k))q^\prime_{ij,\nu}(\bm{x}(t_{k-p})) \label{A6}
  \end{equation}
  Finally we obtain
  \begin{equation}
    \bm{D}_{ij} = \Delta \bm{B}_{ij}\hat{\bm{G}}_{ij}^{-1} \label{A7}
  \end{equation}
  where all elements of vector $\Delta \bm{B}_{ij}$ and maxtrix $\hat{\bm{G}}_{ij}$ can be computed from available data and well defined bases $q_{ij,\mu}$,$q_{ij,\nu}^\prime$ and vector $\bm{D}_{ij}$ can be depicted by Eq.(\ref{A7}).

\begin{acknowledgments}
This work was supported by the National Natural
Science Foundation of China (Grant Nos. 11135001)
and China Postdoctoral Science Foundation (Grant No.2015M581905).
\end{acknowledgments}

\bibliography{basename of .bib file}

\end{document}